\documentclass[final,3p,times,twocolumn]{elsarticle}

\usepackage{lineno}
\usepackage{graphicx}
\usepackage{color}

\usepackage{subcaption}
\usepackage{multirow}
\usepackage{siunitx}

\usepackage{array}
\newcolumntype{L}[1]{>{\raggedright\let\newline\\\arraybackslash\hspace{0pt}}m{#1}}
\newcolumntype{C}[1]{>{\centering\let\newline\\\arraybackslash\hspace{0pt}}m{#1}}
\newcolumntype{R}[1]{>{\raggedleft\let\newline\\\arraybackslash\hspace{0pt}}m{#1}}

\newcommand \gray{$\gamma$-ray}
\newcommand \grays{$\gamma$-rays}

\newcommand \kfz{$^{40}$K}
\newcommand \cosz{$^{60}$Co}
\newcommand \thtTt{$^{232}$Th}
\newcommand \thtTtE{$^{228}$Ra}
\newcommand \thtTtL{$^{224}$Ra}
\newcommand \thtTf{$^{234}$Th}
\newcommand \utTe{$^{238}$U}
\newcommand \utTeE{$^{238}$U}
\newcommand \utTeL{$^{226}$Ra}
\newcommand \utTF{$^{235}$U}
\newcommand \pbtoz{$^{210}$Pb}
\newcommand \potoz{$^{210}$Po}
\newcommand \tltze{$^{208}$Tl}
\newcommand \rnttt{$^{222}$Rn}
\newcommand \geant{\textsc{Geant4}}
\newcommand \dbd{$0\mathrm{\nu\beta\beta}$}

\DeclareSIUnit\year{yr}
\DeclareSIUnit\day{d}
\DeclareSIUnit[number-unit-product = ]\percent{\char`\%}

\DeclareGraphicsExtensions{%
    .pdf,.PDF,%
    .png,.PNG,%
    .jpg,.mps,.jpeg,.jbig2,.jb2,.JPG,.JPEG,.JBIG2,.JB2}

\journal{Astroparticle Physics}

\begin{document}

\begin{frontmatter}

\author[oxf]{P.R.~Scovell\corref{cor1}}
\author[bou]{E.~Meehan\corref{cor2}}
\author[imp]{H.M.~Ara{\'u}jo}
\author[ucl]{J.~Dobson}
\author[ucl]{C.~Ghag}
\author[oxf]{H.~Kraus}
\author[she]{V.A.~Kudryavtsev}
\author[ucl]{X-.R.~Liu}
\author[ral]{P.~Majewski}
\author[bou]{S.M.~Paling}
\author[ral]{R.M.~Preece}
\author[ucl]{R.~Saakyan}
\author[imp]{A.~Tom{\'a}s}
\author[bou]{C.~Toth}
\author[bou]{L.M.~Yeoman}

\address[oxf]{University of Oxford, Department of Physics, Oxford, OX1 3RH, UK}
\address[bou]{STFC Boulby Underground Science Facility, Boulby Mine, Redcar-and-Cleveland, TS13 4UZ, UK}
\address[imp]{Imperial College London, Physics Department, Blackett Laboratory, Prince Consort Road, London SW7 2AZ, UK}
\address[ucl]{University College London (UCL), Department of Physics and Astronomy, Gower Street, London, WC1E 6BT, UK}
\address[she]{University of Sheffield, Department of Physics and Astronomy, Sheffield, S3 7RH, UK}
\address[ral]{STFC Rutherford Appleton Laboratory (RAL), Didcot, OX11 0QX, UK}

\cortext[cor1]{{paul.scovell@stfc.ac.uk}}
\cortext[cor2]{{emma.meehan@stfc.ac.uk}}

\title{Low-background Gamma Spectroscopy at the Boulby Underground Laboratory}

\begin{abstract}

The Boulby Underground Germanium Suite (BUGS) comprises three low-background, high-purity germanium detectors operating in the Boulby Underground Laboratory, located 1.1 km underground in the north-east of England, UK. BUGS utilises three types of detector to facilitate a high-sensitivity, high-throughput radio-assay programme to support the development of rare-event search experiments.  A Broad Energy Germanium (BEGe) detector delivers sensitivity to low-energy gamma-rays such as those emitted by \pbtoz\ and \thtTf. A Small Anode Germanium (SAGe) well-type detector is employed for efficient screening of small samples. Finally, a standard p-type coaxial detector provides fast screening of standard samples. This paper presents the steps used to characterise the performance of these detectors for a variety of sample geometries, including the corrections applied to account for cascade summing effects.  For low-density materials, BUGS is able to radio-assay to specific activities down to \SI{3.6}{\milli\becquerel\per\kilo\gram} for \thtTf\ and \SI{6.6}{\milli\becquerel\per\kilo\gram} for \pbtoz\ both of which have uncovered some significant equilibrium breaks in the \utTe\ chain. In denser materials, where gamma-ray self-absorption increases, sensitivity is demonstrated to specific activities of \SI{0.9}{\milli\becquerel\per\kilo\gram} for \utTeL, \SI{1.1}{\milli\becquerel\per\kilo\gram} for \thtTtE, \SI{0.3}{\milli\becquerel\per\kilo\gram} for \thtTtL, and \SI{8.6}{\milli\becquerel\per\kilo\gram} for \kfz\ with all upper limits at a \SI{90}{\percent} confidence level. These meet the requirements of most screening campaigns presently under way for rare-event search experiments, such as the LUX-ZEPLIN (LZ) dark matter experiment. We also highlight the ability of the BEGe detector to probe the X-ray fluorescence region which can be important to identify the presence of radioisotopes associated with neutron production; this is of particular relevance in experiments sensitive to nuclear recoils.

\end{abstract}



\end{frontmatter}

\section{Introduction}
\label{Sec::Introduction}

The ability to radio-assay materials to ever increasing levels of sensitivity is of great importance to current and next-generation low-background experiments. This is particularly true for detectors located in deep underground laboratories looking for signals from dark matter (DM) or evidence of neutrinoless double beta decay (\dbd). 
In such experiments, primary sources of \gray\ and neutron background radiation come from the intrinsic radioactivity found in the materials from which the detectors are constructed. 
A comprehensive material radio-assay programme to perform careful selection of materials allows this background to be reduced to levels required to meet the experiment's science goals. Moreover, residual activity must be accurately characterised and accounted for in the experiment's background model such that any observed excess may be assessed as potential signal.

Screening programmes typically focus on so-called fixed and mobile contaminants. Fixed contaminants are those which are found embedded in materials and typically consist of naturally occurring radioactive materials (NORM). The most prevalent NORM isotopes are \utTe, \utTF, \thtTt, and their progeny, and the \gray\ emitting isotopes, \kfz, \cosz\ and others. The uranium and thorium chain decays consist of multiple $\alpha$-decays which may produce energetic neutrons through the ($\alpha$,n) reaction and through spontaneous fission. Mobile contamination usually comes primarily from the outgassing of radon from within detector materials which may deposit $\alpha$- and $\beta$-particle emitting progeny on detector surfaces or disperse throughout gaseous or liquid active detector volumes. The characterisation of radon emanation from materials is of particular significance to the low-background community, with direct measurements of radon emanation from materials typically performed in parallel with fixed contamination assays. However, measurement of the \gray\ emitting parent isotope, $^{226}$Ra, can often provide some useful limits for materials whose radon diffusion coefficients and material history are well understood. Generally, more massive materials found close to the active volume of a low-background detector are of the most interest. In the case of noble liquid scintillation detectors, this will comprise items such as photosensors and reflective materials (often PTFE) used to maximise light collection from scintillation. In the case of crystalline detectors, there is particular interest in the intrinsic contamination of the crystals themselves as well as of the materials which form the crystal support structure.

There are several techniques which may be employed to determine the levels of fixed contaminants in materials and one of the most prevalent is through the use of High-Purity Germanium (HPGe) detectors. These detectors are able to spectroscopically determine the levels of NORM isotopes in a material through the detection of the \grays\ associated with their decay. 

The isotope \kfz\ decays through two modes: via $\beta^{-}$-decay to the ground state of $^{40}$Ca with \SI{89.3}{\percent} probability; and via electron capture (EC) nearly always to an excited state of $^{40}$Ar, followed by the emission of a single \gray. This means that the decay of $^{40}$K leads to the emission of a single \SI{1460.8}{\kilo\electronvolt} with a branching ratio of \SI{10.6}{\percent}. The $\beta^{-}$ decay of \cosz\ leads to an excited state of $^{60}$Ni which de-excites via the emission of two \grays\ (\SI{1173.2}{\keV} and \SI{1332.5}{\keV}) in cascade with a branching ratio of almost \SI{100}{\percent}. The decay chains of \utTe, \utTF, and \thtTt\ give rise to the emission of a multitude of \grays. The average number emitted is 2.23, 2.60 and 2.63 per parent decay of the \utTe, \utTF\ and \thtTt\ chains in secular equilibrium, respectively. These numbers include neither conversion electrons (as \grays\ are not emitted) nor X-rays from atomic de-excitation.

\begin{figure}
\centering
  \includegraphics[width=1.0\linewidth]{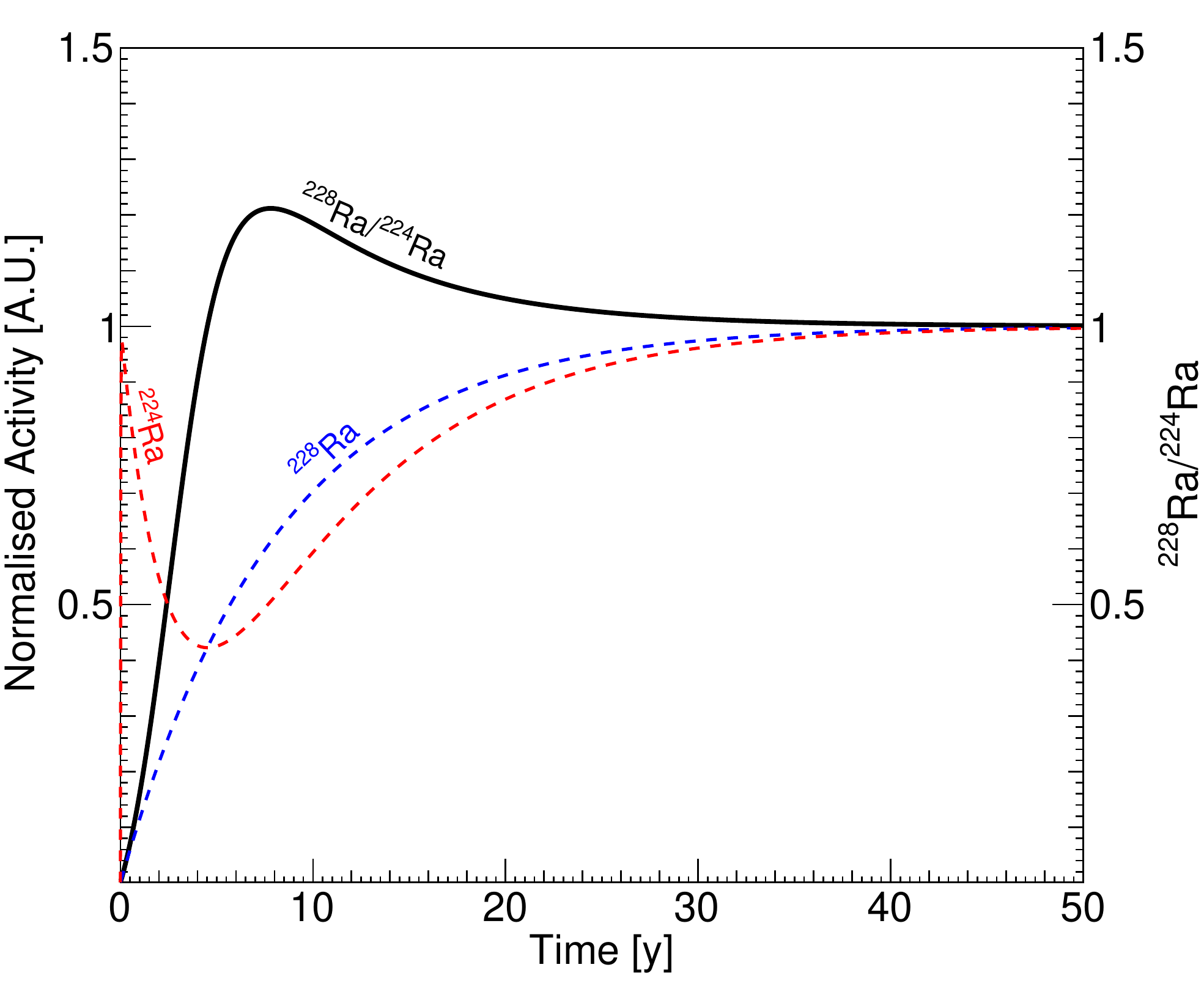}
\caption{Evolution of the normalised specific activity of \thtTtE\ (dashed blue), \thtTtL\ (dashed red), and the evolution of the ratio of the two (thick black) for a material where all radium has been removed. It can be seen that the ratio evolves rapidly over the first 5--10 years after processing (zero time corresponds to the point at which all radium is removed) reaching equilibrium after about 50 years. Additionally, it is only on this timescale that \gray\ spectroscopy will give a true measurement of the specific activity of \thtTt\ in a material such as this. 
\label{fig:bateman}}
\end{figure}

When in secular equilibrium, a measurement of specific activity from any of the \grays\ emitted in the \utTe\ and \thtTt\ quantifies the specific activity of the parent nucleus. However, both chains contain radium, a chemically highly-reactive element, which may be either removed or enriched in material processing. In the case of the \utTe\ chain, this leads to a relatively simple equilibrium break at \utTeL. This means that a measurement of the lower chain isotopes does not necessarily lead to an accurate measurement of the specific activity of \utTe. The \SI{\sim 75}{\kilo\year } half-life of the \utTe\ progeny, $^{230}$Th, means that any break in equilibrium will only recover after many years. This means that the measurement of both \utTeE\ and \utTeL\ may be considered as stable across the life-time of a particle physics experiment. The \thtTt\ chain, however, is not this simple. In this case, we are interested in the ratio of its progeny nuclei \thtTtE\ and \thtTtL\ as these are the isotopes that may be removed or enriched in material processing. If these two isotopes are removed, the chain is re-established  for the early part of the chain through \thtTt\ and for the latter part of the chain through $^{228}$Th. The effect of this is demonstrated in Figure~\ref{fig:bateman} where Bateman equations~\cite{bateman} are used to track the re-establishment of equilibrium in the case where all \thtTt-related radium is removed from a specific material. The half-life of \thtTtE\ (\SI{5.75}{\year}) means that its initial regrowth in the chain is a relatively slow process and it takes some 50 years for it to reach equilibrium with \thtTt. At the same time, the \SI{\sim2}{\year} half-life of $^{228}$Th means that measured specific activity for \thtTtL\ will decrease over a period of about 5 years. At this point, the ratio of \thtTtE:\thtTtL\ reaches unity but again, the half-life of $^{228}$Th means that the specific activity of \thtTtE\ then becomes greater than that of \thtTtL. After a period of \SI{\sim50}{\year} equilibrium is reached and the full chain is representative of the specific activity of \thtTt. In \gray\ spectroscopy, one is not able to measure \thtTt\ directly and must instead rely on \grays\ from later in the chain. In the case where \thtTtE\ and \thtTtL\ are measured in equilibrium, it can be assumed that they represent the true specific activity of \thtTt.

\textcolor{black}{In the case of the \utTe\ chain, it is possible that equilibrium may also be broken through the process of emanation of \rnttt. The emanation of \rnttt\ will not ordinarily cause time dependent change in the apparent  \utTeE:\utTeL\ ratio but it is known that the radon diffusion coefficient has a strong dependence on temperature with the probability of emanation falling with decreasing temperature~\cite{ISKANDAR2004971}. It is important, therefore, that a comprehensive materials radio-assay campaign includes measurements of radon emanation; ideally at the temperature at which materials will be held during detector operation.}

To establish if any equilibrium breaks are present in these chains, the typical energy range of \grays\ of interest ranges from 46.5~keV (from the decay of \pbtoz) to \SI{2614.5}{\keV} (from \tltze). However, we will show in this paper that there is important information to be gained from the study of X-ray energies down to at least \SI{15}{\kilo\electronvolt}. The energy range of these \grays\ suggests that a screening programme may only be considered truly comprehensive when the HPGe detectors used are sensitive over the same energy range. Inductively-coupled plasma mass spectrometry (ICP-MS) may reach better sensitivity to material contaminations of \utTe\ and \thtTt\ but this is a destructive technique and it is currently not possible to study any deviation from equilibrium of progeny in each chain.

\section{Boulby Underground Laboratory}
\label{Sec::Boulby}

The Boulby Underground Laboratory is located at a depth of \SI{1100}{\meter} (2,840 metres water equivalent) and is operated by the UK's Science and Technology Facilities Council. It has a rich and significant history in the development and support of low-background physics. Until recently, the main experimental area underground was the Palmer Laboratory which hosted the pioneering ZEPLIN dark matter programme that operated a series of three xenon detectors until 2011~\cite{Alner:2005pa,Alner:2007ja,Akimov:2006qw,Lebedenko:2008gb}. Similarly, the leading directional dark matter programme, DRIFT, was hosted in this laboratory~\cite{Battat:2016xxe}. Both ZEPLIN-I and, additionally, the NAIAD dark matter experiment~\cite{SPOONER2000330} were hosted in other areas of the mine.

\begin{figure}
\centering
\subcaptionbox*{}{%
  \includegraphics[width=1.0\linewidth]{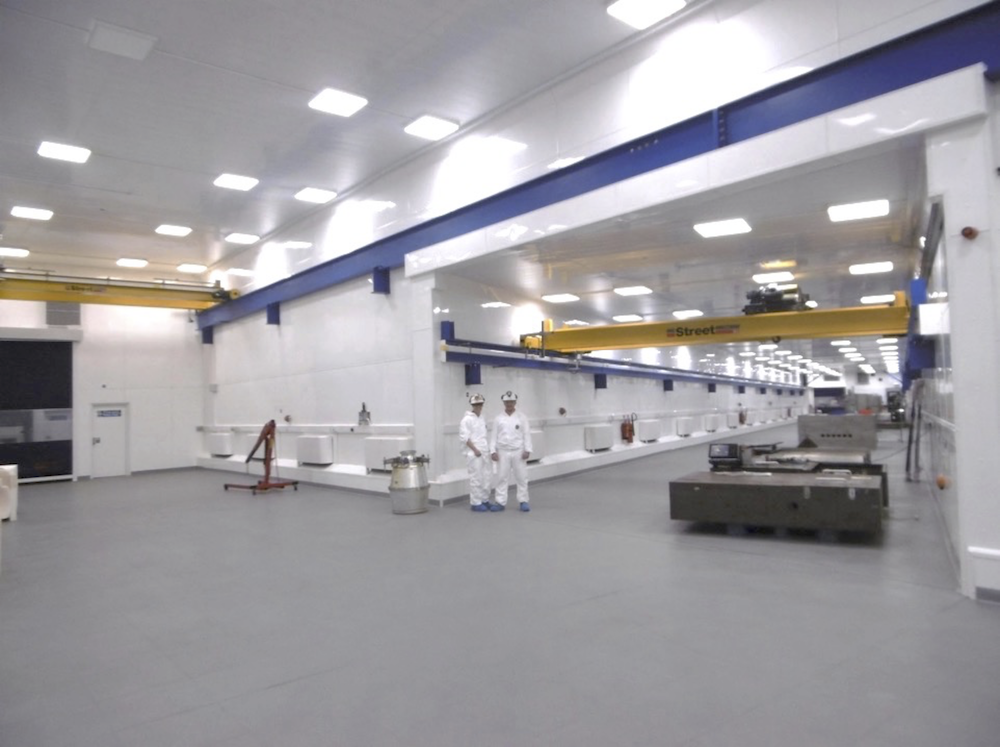}
  }\vspace{-1.em}
\subcaptionbox*{}{%
  \includegraphics[width=1.0\linewidth]{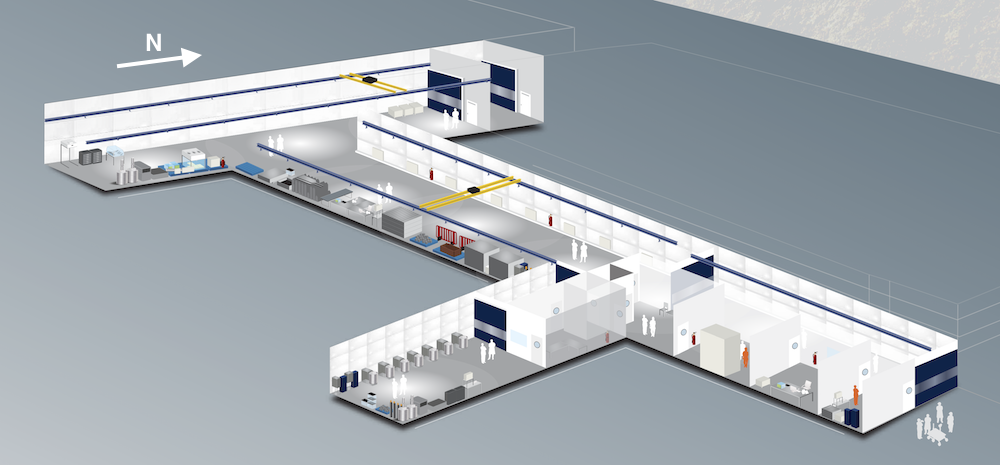}
}
\caption{(top) A picture taken from the Large Experimental Cavern down the length of the new Boulby Underground Laboratory. (bottom) A CAD rendering of the new laboratory. The dedicated area for low-background screening is situated half way along the main laboratory and the Large Experimental Cavern runs north (labelled $N$) to south at the top of the rendering.
\label{fig:thelab}}
\end{figure}

In 2015 a new laboratory area was constructed to replace the now decommissioned Palmer Laboratory that was approaching the end of its life. The new laboratory, shown in Figure~\ref{fig:thelab}, is an improvement in terms of infrastructure, volume, and cleanliness. Unlike the wood and plasterboard frame of the Palmer Laboratory, the new facility is of a more robust steel-based design which will maintain the integrity of cleanliness and provides a degree of radiation shielding to the rock which forms the experimental cavern. It consists of \SI{4000}{\cubic\metre} of experimental space, with a \SI{4}{\meter} high and \SI{7}{\meter} wide main hall. A connected area known as the Large Experimental Cavern (LEC) has a height of \SI{6.5}{\meter}. These areas of the laboratory are certified to ISO class 7 (10k) cleanroom standard and are serviced by $x$-$y$ cranes that facilitate simple transfer of materials around the lab. A further \SI{15.5}{\meter} long and \SI{7}{\meter} wide area is maintained at ISO Class 6 (1k) cleanroom standard and is dedicated to low-background counting. The geology of the cavern rock around the Boulby Underground Laboratory contributes its suitability for low-background activities. The halite rock has been measured to contain \SI{32(3)}{ppb} of \utTe, \SI{160(20)}{ppb} of \thtTt, and \SI{0.036(3)}{\percent} of potassium~\cite{Malczewski2013}. All \grays\ originating in the cavern rock are effectively attenuated through the use of detector shielding. However, most significantly, the low level of \utTe\ contributes to a low ambient background from airborne $^{222}$Rn of only \SI{2.4}{\becquerel\per\cubic\meter}~\cite{Araujo:2011as}. This is significantly lower than the lowest radon activity values measured in the Gran Sasso laboratory of \SIrange{20}{50}{\becquerel\per\cubic\meter}\cite{Arpesella:1994ew}.

\section{Boulby Underground Germanium Suite (BUGS)}
\label{Sec::BUGS}

\begin{figure}
\centering
  \includegraphics[width=1.0\linewidth]{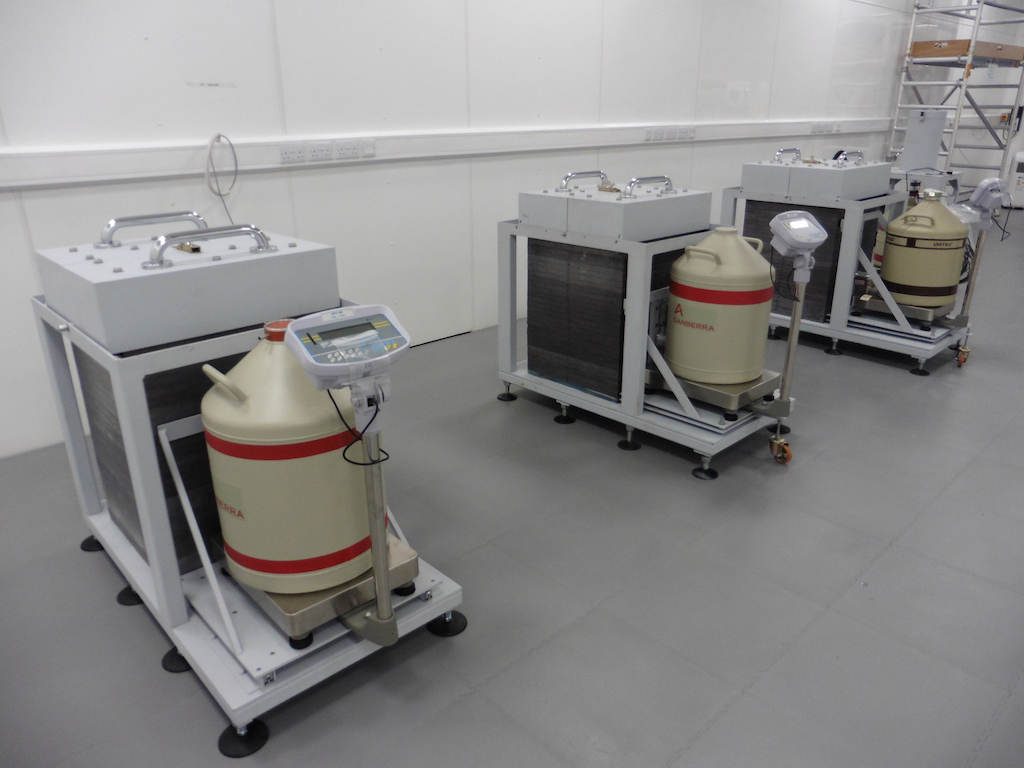}
\caption{(l-r) Lumpsey, Chaloner, and Lunehead, the three detectors of BUGS. The small pre-screening detector (Wilton) can be seen to the right of Lunehead in a taller, cylindrical shield.
\label{fig:AllDetectors}}
\end{figure}

In order to redevelop the capability for radioactivity screening in the UK and provide support for rare-event search experiments with major UK involvement, a new facility has been developed, primarily by a collaboration between the Boulby Laboratory and the DMUK consortium (a collaboration between UK scientists involved in dark matter search experiments).
The facility, known as `BUGS' (Boulby Underground Germanium Suite), is housed in the ISO Class 6 cleanroom low-background counting suite and includes three primary ultra-low-background HPGe- detectors: Chaloner, Lunehead and Lumpsey (shown \textit{in situ} in Figure~\ref{fig:AllDetectors}), and a fourth small detector, Wilton, operated only as a pre-screening device \textcolor{black}{to give qualitative measurements of materials suspected to be of high activity. Materials that give measurable lines in Wilton will generally be of too high activity for use in low-background experiments and, as such are not screened in our more sensitive detectors.} Since 2015, BUGS has supported primarily the LUX-ZEPLIN\footnote{A collaboration between members of the former LUX and ZEPLIN experiments} (LZ) construction material radio-assay campaign~\cite{Akerib:2015cja,Mount:2017qzi,Akerib:2017iwt}, as well as performing assays for the SuperNEMO \dbd\ experiment~\cite{Piquemal:2006cd} and the Super-Kamiokande experiment~\cite{MORI2013316}. The current capacity of BUGS allows for $\sim$100 sample assays per year \textcolor{black}{with some samples assayed in a matter of days and others requiring several weeks to reach the required sensitivity}. BUGS will contribute to the radio-assay campaigns of several other experiments with UK research interests in the near future, in addition to providing support for the industrial and commercial sectors.

\begin{figure}
\centering
\subcaptionbox*{}{%
  \includegraphics[width=1.0\linewidth]{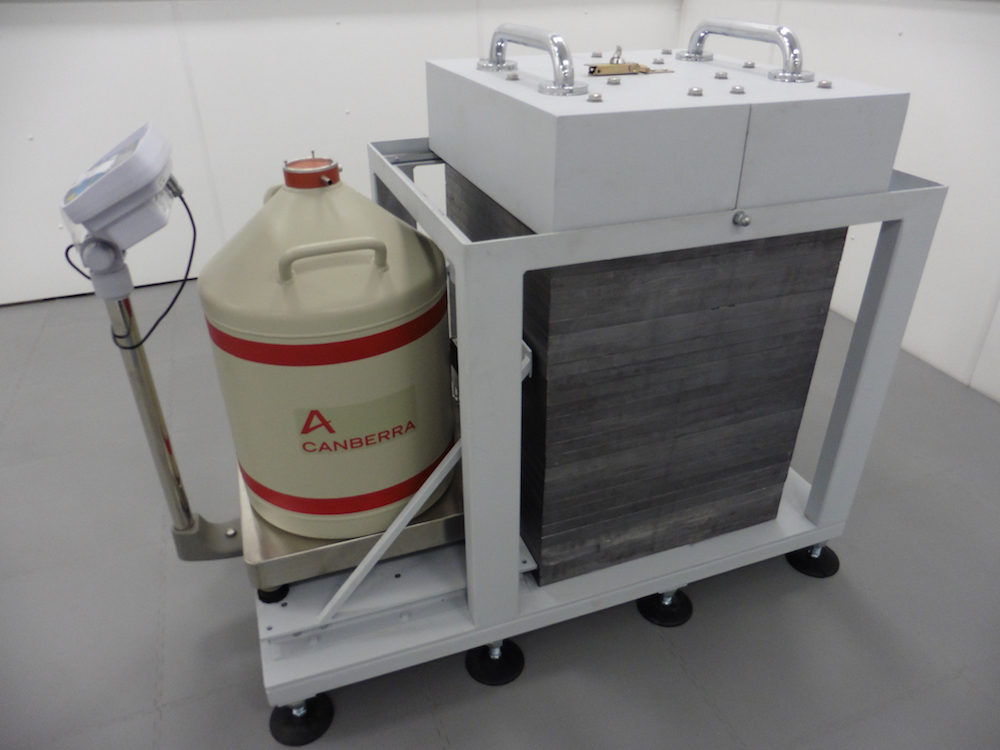}
  }\vspace{-1.em}
\subcaptionbox*{}{%
  \includegraphics[width=1.0\linewidth]{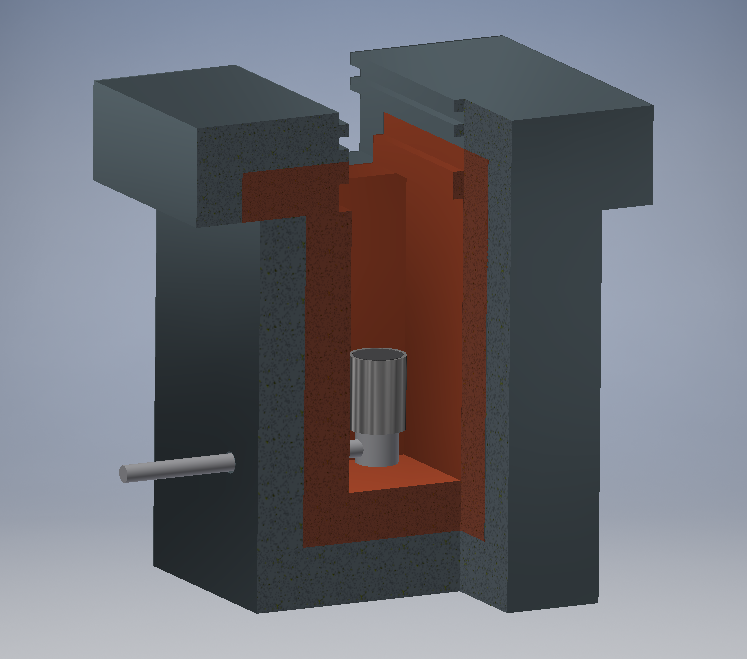}
}
\caption{(top-bottom) An image and CAD rendering of the shield used on one of the BUGS detectors. In the image, the liquid nitrogen dewar can be clearly seen sitting on a scale which is used to monitor the remaining liquid. The lid is retracted by pulling the two stainless steel handles on top. The CAD image shows the multi-layer format of the shield and the j-type neck used on all of the BUGS detectors. Also visible are the chevron edges of the lid section which prevent line of sight from the inside to the outside of the shield. The stainless steel frame of the shield is not reproduced in the CAD representation.
\label{fig:shield}}
\end{figure}

The three BUGS detectors are housed in shielding castles developed in collaboration with Lead Shield Engineering Ltd~\cite{leadshield}. In all cases, the shields are constructed using \SI{9}{\cm} thickness of lead outside \SI{9}{\cm} thick OHFC copper. The copper and lead come from Boulby stock material that has remained underground for at least 2 decades, and selected for the BUGS castles after \textcolor{black}{qualitative assay of batches to determine those with the lowest radioactivity}. The internal cavity of each castle is of dimensions 
\SI[product-units = brackets]{200x220x510}{\milli\meter\cubed}, allowing for the radio-assay of large samples. The varying geometries of the HPGe detector cryostats mean that in each case the distance between the detector face and the bottom of the shield lid varies. For Chaloner, it is \SI{29}{\cm}, for Lunehead \SI{16}{\cm}, and for Lumpsey \SI{27}{\cm}. The three shields are purged with N$_{2}$ gas to remove air-borne radon in the shield cavity. The N$_{2}$ gas purge line enters the castle through a meandering bore in the shielding to minimise the risk of increasing the detector background through a line of sight aperture to the outside. The shields include inter-locking retractable roofs in order to simplify sample loading. An image of one of these shields alongside a cut-through CAD model is shown in Figure~\ref{fig:shield}. The detector dewars sit on scales that monitor the weight of liquid N$_{2}$ remaining for cooling. Both the liquid and gas N$_{2}$ are provided by a model LN65 nitrogen liquefier from Noblegen which includes a \SI{300}{\litre} storage dewar. The LN65 can produce \SI{65}{\litre} of liquid N$_{2}$ per day which adequately provides for the detectors of BUGS. The LN65 also includes a pressure swing adsorption nitrogen gas generator which provides \SI{3}{\litre\per\minute} of gaseous N$_{2}$ to each of the castles.


\begin{figure}
\centering
\subcaptionbox*{}{%
  \includegraphics[width=1.0\linewidth]{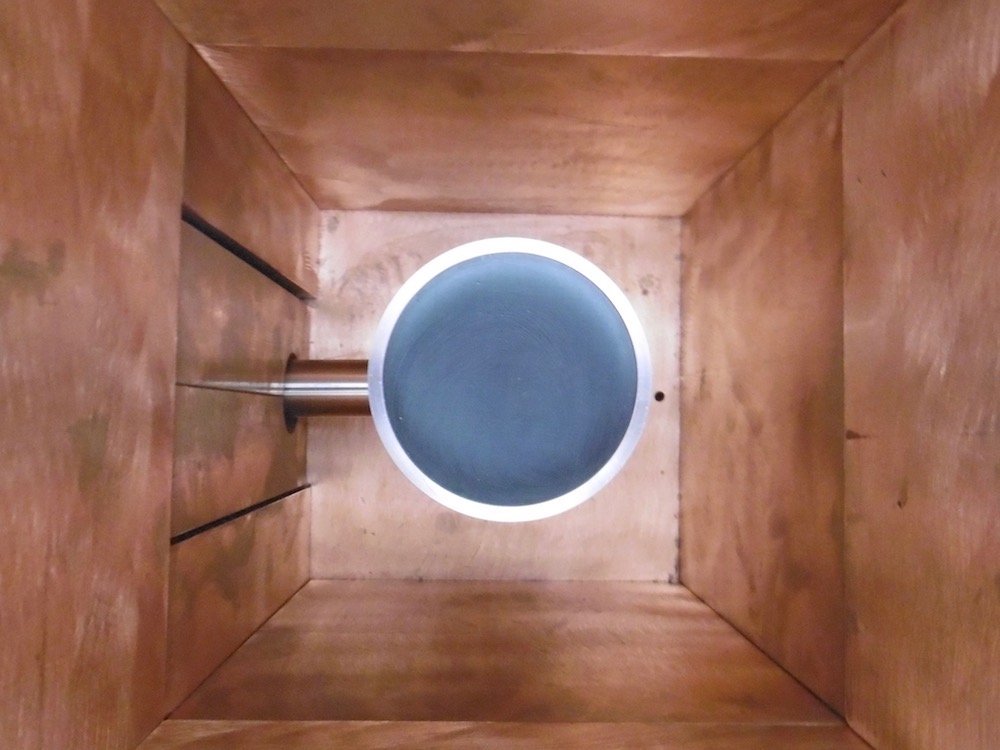}
  }\vspace{-1.em}
\subcaptionbox*{}{%
  \includegraphics[width=1.0\linewidth]{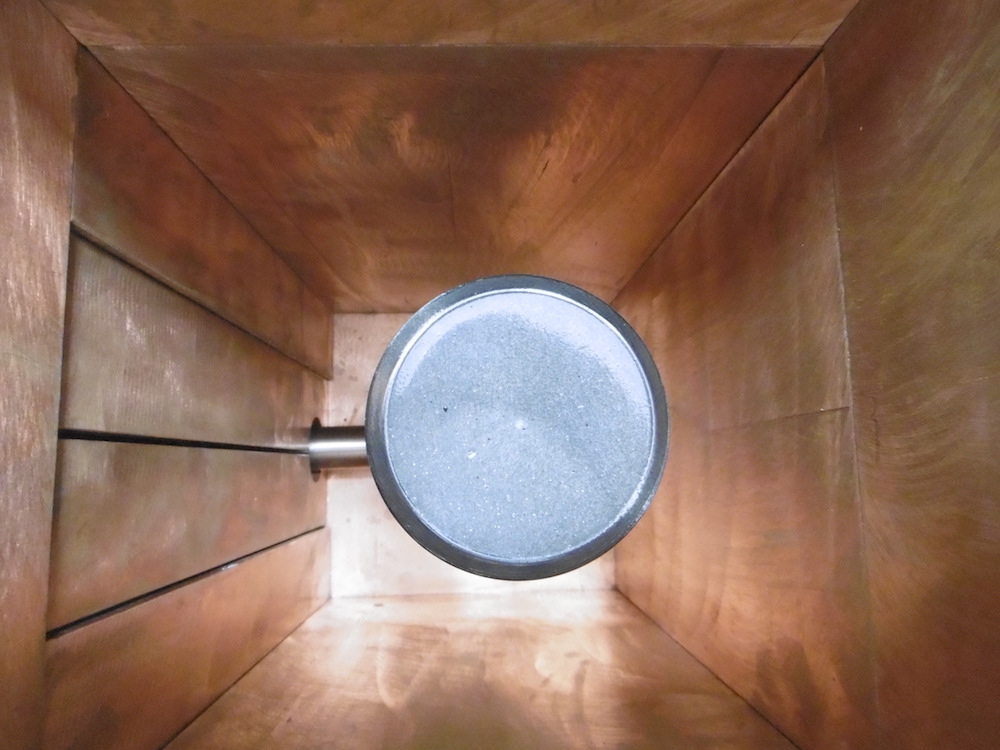}
  }\vspace{-1.em}
  \subcaptionbox*{}{%
  \includegraphics[width=1.0\linewidth]{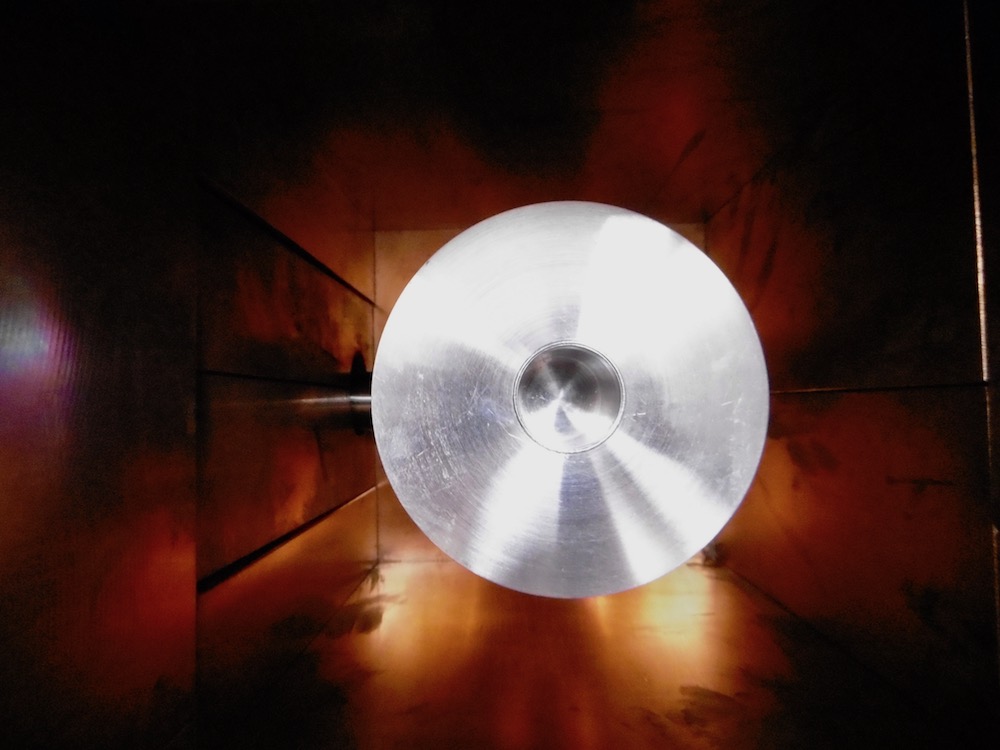}
  }  
\caption{Images of the three BUGS detectors. The Chaloner BEGe detector with an end cap diameter of \SI{4}{inches} (top). The Lunehead p-type coaxial HPGe detector with an end cap diameter of \SI{3.75}{inches} (middle). The Lumpsey SAGe-well detector with an end cap diameter of \SI{4.25}{inches} (bottom).
\label{fig:detectors}}
\end{figure}

BUGS deploys a range of HPGe detector types rather than using only standard coaxial HPGe detectors. We have chosen a broad range of detector types in order to probe the maximum number of possible sample geometries and to provide high sensitivity to the full \gray\ energy range discussed above. BUGS currently represents one of the most comprehensive germanium screening facilities for low-background physics worldwide. The three primary detectors are shown in Figure~\ref{fig:detectors}. The next sections of this paper discuss each of the detectors in turn with emphasis on what makes each particularly suitable for materials assay, software developed for BUGS, the detector characterisations and, finally, sensitivity to typical material samples expected from low-background experiments.

\subsection{Chaloner (BEGe)}

Chaloner is a \SI{\sim0.8}{\kilogram} BE5030 Broad Energy Germanium (BEGe) detector manufactured by Mirion (Canberra). This detector has a nominal front face surface area of \SI{50}{\centi\meter\squared}, a length of \SI{30}{\milli\meter} and a measured relative efficiency of \SI{48}{\percent} (defined as relative to the efficiency to \SI{1332}{\kilo\electronvolt} $^{60}$Co \grays\ of a \SI{3x3}{inch} NaI detector with a source-detector distance of \SI{25}{\centi\meter}). The BEGe detector is constructed in a planar format and has a small inner electrode for signal output. The geometry of this detector allows for an effectively dead-layer free front face which gives excellent efficiency for low-energy \grays\ such as the \SI{46.5}{\kilo\electronvolt} line associated with the $\beta$-decay of $^{210}$Pb to $^{210}$Bi in the \utTe\ series. In addition to high efficiency to low-energy \grays, a highly selective crystal choice (the crystal doping profile is selected to give improved charge collection) allows for greater peak resolution and improved peak shape at higher energies. This detector allows the screening of the full radium series --- previously not possible for the screening of NORM isotopes in material searches for low-background detectors. The detector cryostat includes a carbon fibre end-cap to allow for the maximum transmission of low-energy \grays, shown in the upper image of Figure~\ref{fig:detectors}.


\subsection{Lumpsey (SAGe well)}

\begin{figure}
\centering
  \includegraphics[width=1.0\linewidth]{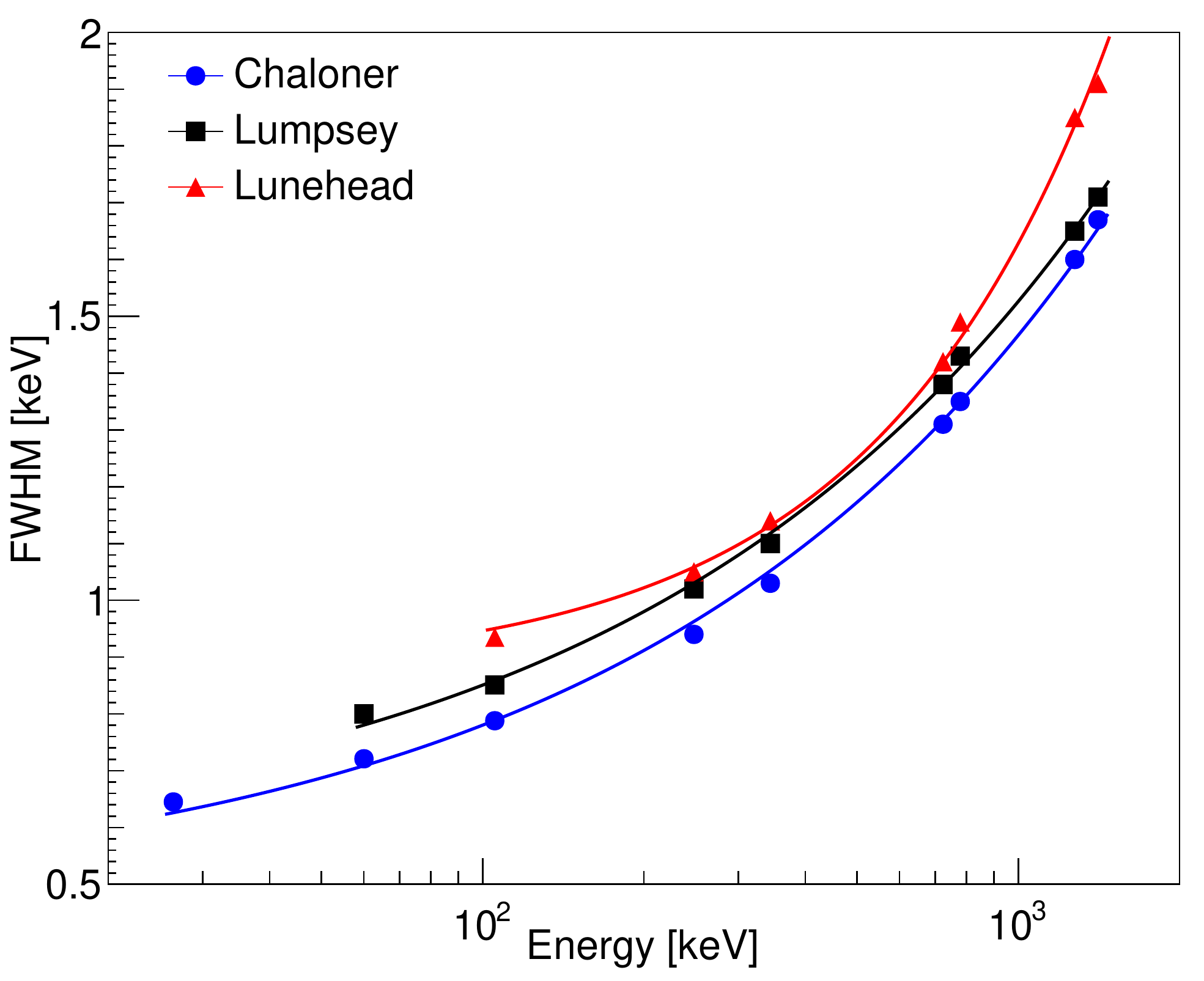}
\caption{Measured resolution for the three BUGS detectors. Lunehead (red) shows a resolution comparable to Lumpsey (black) for the \SIrange{400}{700}{\kilo\electronvolt} energy range. The resolution in Lunehead is superior to that seen in larger co-axial p-type detectors.
\label{fig:DetectorResn}}
\end{figure}

Lumpsey is a \SI{1.5}{\kg} Mirion (Canberra) GSW275L small anode germanium (SAGe) well-type detector with a nominal minimum detector volume of \SI{275}{\cm\cubed} and a well diameter and depth of \SI{28}{\mm} and \SI{40}{\mm}, respectively. Lumpsey has a measured relative efficiency of \SI{69}{\percent}. The geometry of the detector allows almost $4\pi$ coverage for samples screened in the well thus allowing high efficiency screening for small samples. As with the BEGe detector, the SAGe well detector allows excellent efficiency to low-energy \grays. Previous generations of well detector have suffered from poor resolution but the SAGe well detector gives a response approaching that of the BEGe detector --- as shown in Figure~\ref{fig:DetectorResn}. The crystal is held in a stainless-steel cryostat. The Lumpsey detector has the added advantage that it can be used as a standard coaxial HPGe detector for larger samples placed outside of the well.

\subsection{Lunehead (p-type coaxial)}

\begin{figure}
\centering
  \includegraphics[width=1.0\linewidth]{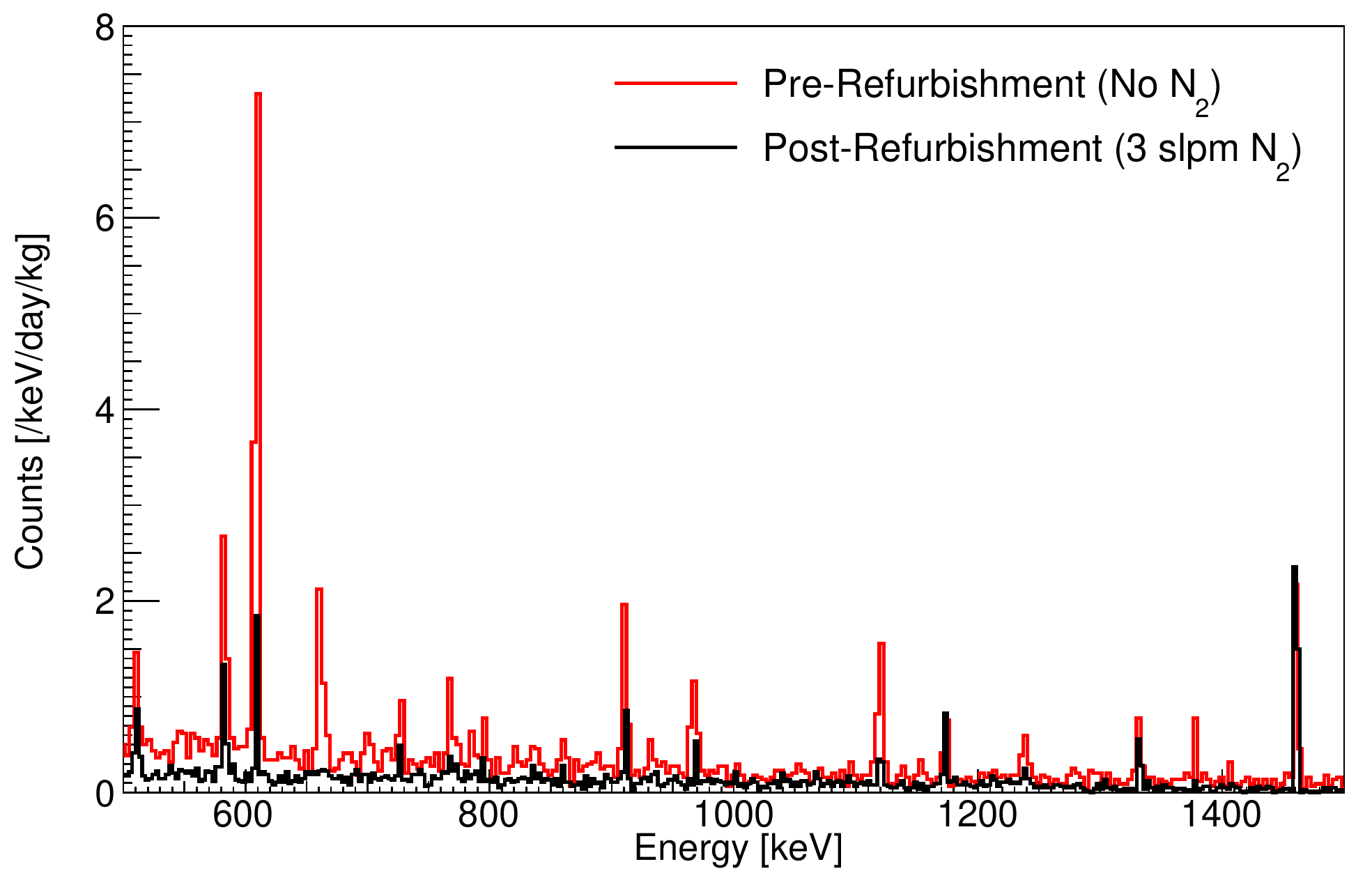}
\caption{Improvement to the background of the Lunehead detector post-refurbishment (black) compared to its previous configuration and shielding (red). In addition to the carbon fibre end-cap, the detector is now housed in new shielding which is flushed using \SI{3}{\litre\per\minute} of gas N$_{2}$ to reduce background due to air-borne $^{222}$Rn. In both cases we see an identical \kfz\ background \textcolor{black}{rate} of $\sim$\SI{210}{\micro\becquerel}.
\label{fig:lunhead_improv}}
\end{figure}

Lunehead is a \SI{2.0}{\kg} Ortec GEM-XX240-S standard p-type coaxial detector with a nominal relative efficiency of \SI{92}{\percent}~\cite{ortec}. This detector was previously used to screen materials for the ZEPLIN and DRIFT programmes. 
In order to improve the background and low-energy sensitivity substantially, the detector was returned to the manufacture for refurbishment. The magnesium end-cap was replaced with one of ultra-low-background carbon fibre. This both reduces the intrinsic background of the detector and increases efficiency to lower energy \grays. Figure \ref{fig:lunhead_improv} shows the improvement achieved following refurbishment. The integrated count rate for \SI{>100}{\keV} decreased from 4.2 to 0.8 counts per minute. The background for sub-\rnttt\ \grays\ is reduced below the ambient level in the BUGS laboratory due to the N$_{2}$ purge through the lead and copper shielding. 

\section{Detector Characterisation}

\subsection{Acquiring Spectra and Data Analysis}
\begin{figure}
\centering
  \includegraphics[width=1.0\linewidth]{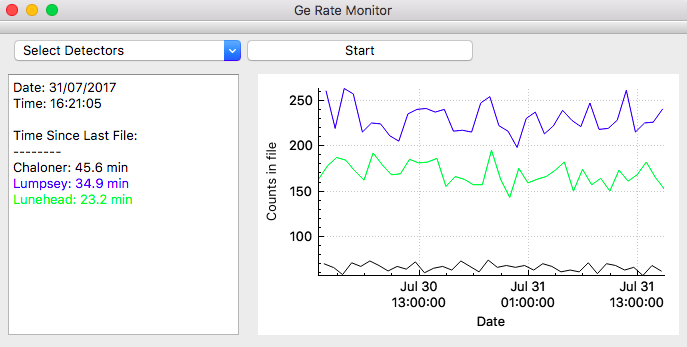}
\caption{The GeMonitor software looking at data from the three detectors. The software allows users to monitor both count rates and the time since the last file was acquired. If the time goes significantly over an hour, operators are informed via instant messenger.
\label{fig:GeMonitor}}
\end{figure}

Lunehead data acquisition is performed using an ORTEC DSPEC-50 single Multi-Channel Analyser (MCA), whilst both Chaloner and Lumpsey data are acquired using Mirion LNYX Digital Signal Analysers (DSAs). To monitor detector stability, the acquisition is split into 1-hour long files. Data output and stability are monitored using custom software, GeMonitor, developed using QT~\cite{qt} and incorporating the QCustomPlot~\cite{qcustomplot} libraries. GeMonitor will automatically contact BUGS collaborators if acquisition has ceased on any of the detectors. Figure~\ref{fig:GeMonitor} shows a screenshot of the GeMonitor software looking at the Chaloner detector. The user is able to select the look-back time and apply a threshold. 

In order to facilitate consistent analysis of data from the BUGS detectors, a standalone piece of analysis software has been produced. The BUGS Analysis Software Suite (BASS) has been developed in QT and incorporates ROOT~\cite{root}, qcustomplot, and QZXing~\cite{qzxing} libraries. BASS allows the user to open data files in several known formats and can combine multiple files to form a single spectrum. Once opened, the user is easily able to identify and automatically remove any files within which an unexpected rate is seen and may inspect individual files if this is required. A common cause of elevated rate is from acoustically induced noise due to the filling of the LN$_{2}$ dewars in the Lunehead and Lumpsey detectors. \textcolor{black}{In Lunehead, we see this rate increase in the \SIrange{0}{100}{\keV} energy range and in Lumpsey, we see this in the \SIrange{0}{200}{\keV} energy range. For Chaloner, the effect is only seen below \SI{10}{\keV} so does not impact quantitative measurement in the energy range of interest.} BASS is able to read in full photopeak efficiency (discussed in Section~\ref{Sec::Simulation}) files in both ROOT and plain text format and can combine integrated line rates with these and a user-defined sample mass to output specific activities for any given line. A library of standard decays provides the \gray\ intensities needed for analysis of common NORM isotopes. The user is able to output a PDF report which presents all results in a consistent format. Custom details of detector names and background rates (used in the determination of specific activity) may be easily added to the software by the user. BASS has been benchmarked against results from Gammavision, Gamma Acquisition \& Analysis and the PeakEasy~\cite{peakeasy} software.

\begin{figure}
\centering
  \includegraphics[width=1.0\linewidth]{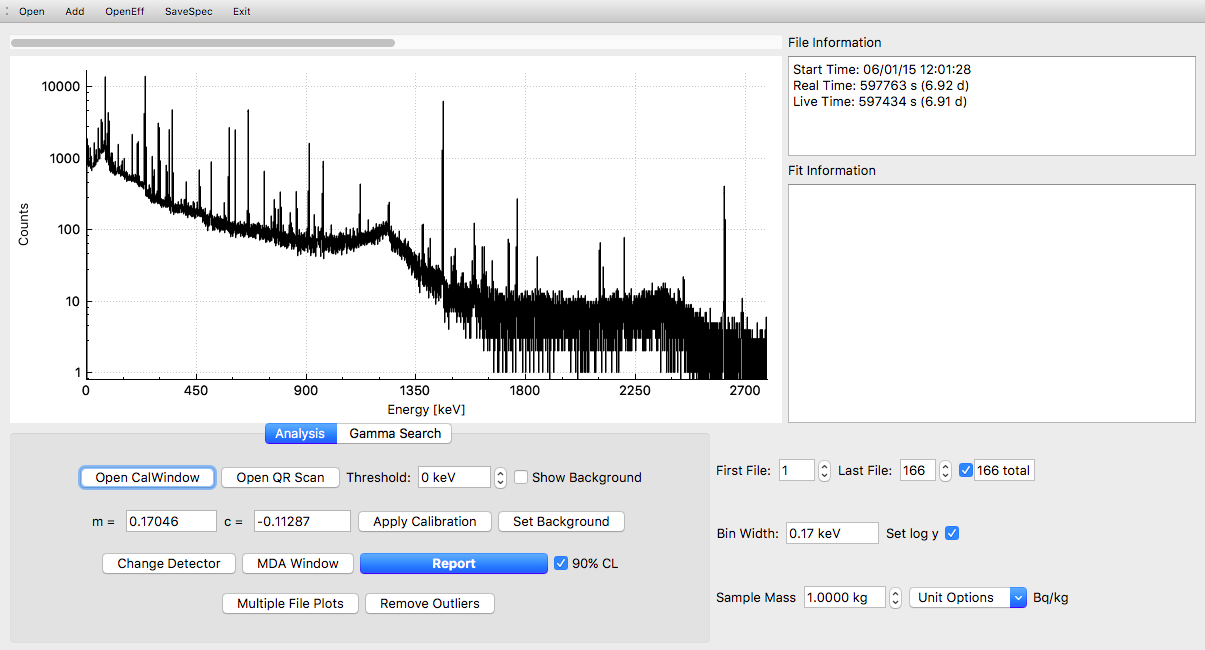}
\caption{BASS looking at an IAEA385 standard calibration spectrum acquired using Chaloner.
\label{fig:BASS}}
\end{figure}

In addition to basic analysis, BASS includes a peak identification algorithm which allows the user to click on a point of interest and see candidate decay lines. The candidate lines may be ordered either by intensity, parent isotope, or by the magnitude of energy difference from the point of interest. This has proven to be invaluable in the identification of unexpected decay lines. BASS is also able to calculate minimum-detectable activity curves (discussed in Section~\ref{sec:mds}) for any given sample. This aids the user in determining the length of run needed to reach the required sensitivity for a sample. Figure~\ref{fig:BASS} shows a screenshot of BASS where the various user inputs may be seen. Finally, BASS also includes the ability to scan 2d barcodes such as those used for sample identification and cataloging. The software scans an item and will open the embedded URL in the user's default web browser. This greatly simplifies the task of matching radio-assay reports to materials or components.

\subsection{Simulating the Full Photopeak Efficiency}
\label{Sec::Simulation}

\begin{figure}
  \begin{subfigure}[b]{0.15\textwidth}
    \includegraphics[width=\textwidth]{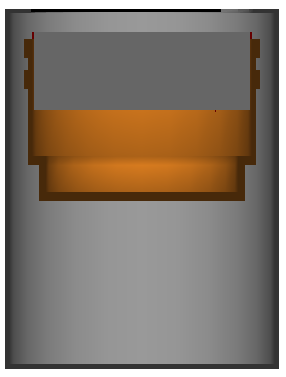}
  \end{subfigure}
  \begin{subfigure}[b]{0.15\textwidth}
    \includegraphics[width=\textwidth]{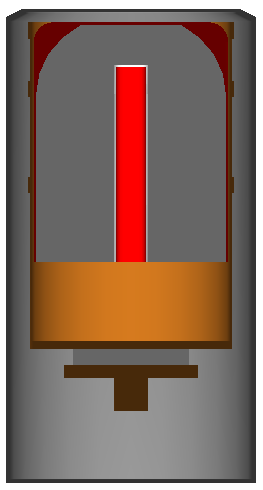}
  \end{subfigure}
  \begin{subfigure}[b]{0.15\textwidth}
    \includegraphics[width=\textwidth]{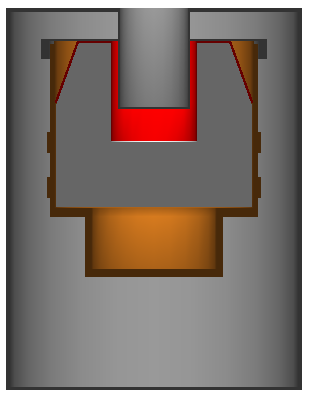}
  \end{subfigure}
  \caption{\geant\ renderings of the three principle BUGS detectors. (left) The Chaloner BEGe detector. (middle) The Lunehead p-type coaxial HPGe detector. (right) The Lumpsey SAGe-well detector. The germanium crystals are represented in \textcolor{black}{grey} with the red regions representative the characterised dead-layer profile for each. It can be seen that Lunehead has substantial inactive volumes towards the front face of the detector. Each germanium crystal is held in a copper holder and is surrounded by its cryostat.
\label{fig:detectorsG4}}
\end{figure}

In order to accurately predict the sample-specific full photopeak efficiencies for each detector, a simulation package has been developed using the \geant.10.2 toolkit~\cite{Agostinelli:2002hh} that describes the Ge crystal, detector housings including the main internal components, and the detector shields. Detector cryostat and crystal geometries are provided by the manufactures to construct a detector model, refined through our own calibrations and characterisations of the instruments. Figure~\ref{fig:detectorsG4} shows the \geant\ renderings of the three detectors.

The basic methodology for determining the full photopeak efficiency of any sample is as follows. Firstly, the sample must be translated into a \geant\ physical volume. Sample materials (including containers and powders) are described using the \geant\ table of NIST materials or are described with user-defined materials where this is not possible. 
With detector geometry defined, it is possible to calculate the full photopeak efficiency for any given sample. This is achieved by first defining a representative sample geometry, henceforth referred to as the sample volume. A flat spectrum of \SIrange{0}{3}{\MeV} \grays\ is released uniformly and isotropically from the sample volume with any energy deposited in the detector crystal recorded. \textcolor{black}{To allow for the finite resolution of any HPGe detector,} efficiency is then calculated by measuring the fraction of \grays\ which have deposited over \SI{95}{\percent} of their initial energy in the detector as a function of energy. 
The efficiency depends on the composition, shape, and location of each sample and must be individually calculated.
Aside from the size of the detector crystal, which may be accurately measured by Canberra, the two factors which dominate the response of the detector are the detector dead-layer thickness and the distance between the detector crystal and the cryostat window. It is common to use boron implantation to create a p-type contact and lithium diffusion to create an n-type contact. The thickness of the contact constitutes an inactive layer in the detector, more commonly referred to as a dead-layer. The construction of the BEGe detector uses a proprietary method to produce a contact which leads to an almost zero thickness dead-layer on the front face. Modifying the dead-layer thickness in simulation gives rise to an energy dependent variation in calculated efficiency. Efficiency drops off rapidly at lower energy as the dead-layer thickness increases as lower energy \grays\ (below $\sim$\SI{100}{\kilo\electronvolt}) are more strongly attenuated. Higher energy \grays\ only lose efficiency due to the overall reduction in active detector mass caused by the increase in dead-layer thickness. Modifying the crystal-window distance gives rise to an energy independent modification of detector efficiency. This is simply due to the fact that a larger distance means that a smaller solid angle is subtended by the detector as the effective distance between sample and crystal increases. This effect is not seen when a Marinelli type beaker (in which the sample is placed both on the front and around the sides of the detector crystal) is used. In screening programmes for low-background particle physics detectors, it is usual to receive modestly sized samples which may not be suitable for placement in Marinelli beakers and instead will be placed on the face of a detector. For samples such as these, it is important to determine the correct crystal-sample distance.
 
Initial characterisation was performed using a multi-gamma source. In the case of Chaloner, the dead-layer thickness was probed using the relative peak heights of \SI{86.5}{\keV} and \SI{105.3}{\keV} \grays\ from $^{155}$Eu decay. As Lunehead and Lumpsey are insensitive to low-energy \grays, the same characterisation was performed using \SI{121.8}{\keV} and \SI{244.7}{\keV} \grays\ from $^{152}$Eu decay. These decays are simulated and the dead-layer thickness modified until simulation accurately matches data with no relative scaling between peaks. This gives a baseline characterisation which we can adjust using a source which is more representative of what will be screened on these detectors.

A typical sample material screened for a low-background physics experiment will not be a point source. This being the case, the detectors must be characterised for their response to extended sources placed on the detector face. A suitable material for constructing calibration geometries has been found to be IAEA385 powder~\cite{IAEA385}. IAEA385 is a standard calibration material, derived from a \SI{250}{\kg} sample of Irish Sea sediment collected in 1995. This sample was analysed with a variety of methods at 99 independent laboratories to give a calibrated table of isotope concentrations. In simulation, we describe the IAEA385 material using the elemental stoichiometry: O50:Si27:Al7:Ca4:Fe4:K2:Mg2:C2:Na1 and a density of \SI{1.22}{\gram\per\cubic\centi\meter}. To characterise each detector a sample was produced housed in a \SI{1}{inch} diameter pot filled to a level of \SI{50}{\mm}. For characterisation of coaxial detectors, this is not the most optimal configuration but, in the case of cross-calibration, it is useful to measure an identical sample across all detectors.

\subsection{True Coincidence Summing Corrections}
\label{sec:CCF}

In all detectors, coincidence summing effects are important. In the case of a well detector, a sample placed in the well may have almost 4$\pi$ coverage. This being the case, it is far more likely that multiple \grays\ emitted within the acquisition timing window of the detector will deposit some or all of their energy in the crystal. We must include true coincidence summing corrections in our detector characterisations. The method used to calculate these corrections for a SAGe well detector similar to the one operated by BUGS is described in depth in~\cite{Britton:2015aa} and only briefly summarised here. 

\begin{figure}
\centering
  \includegraphics[width=1.0\linewidth]{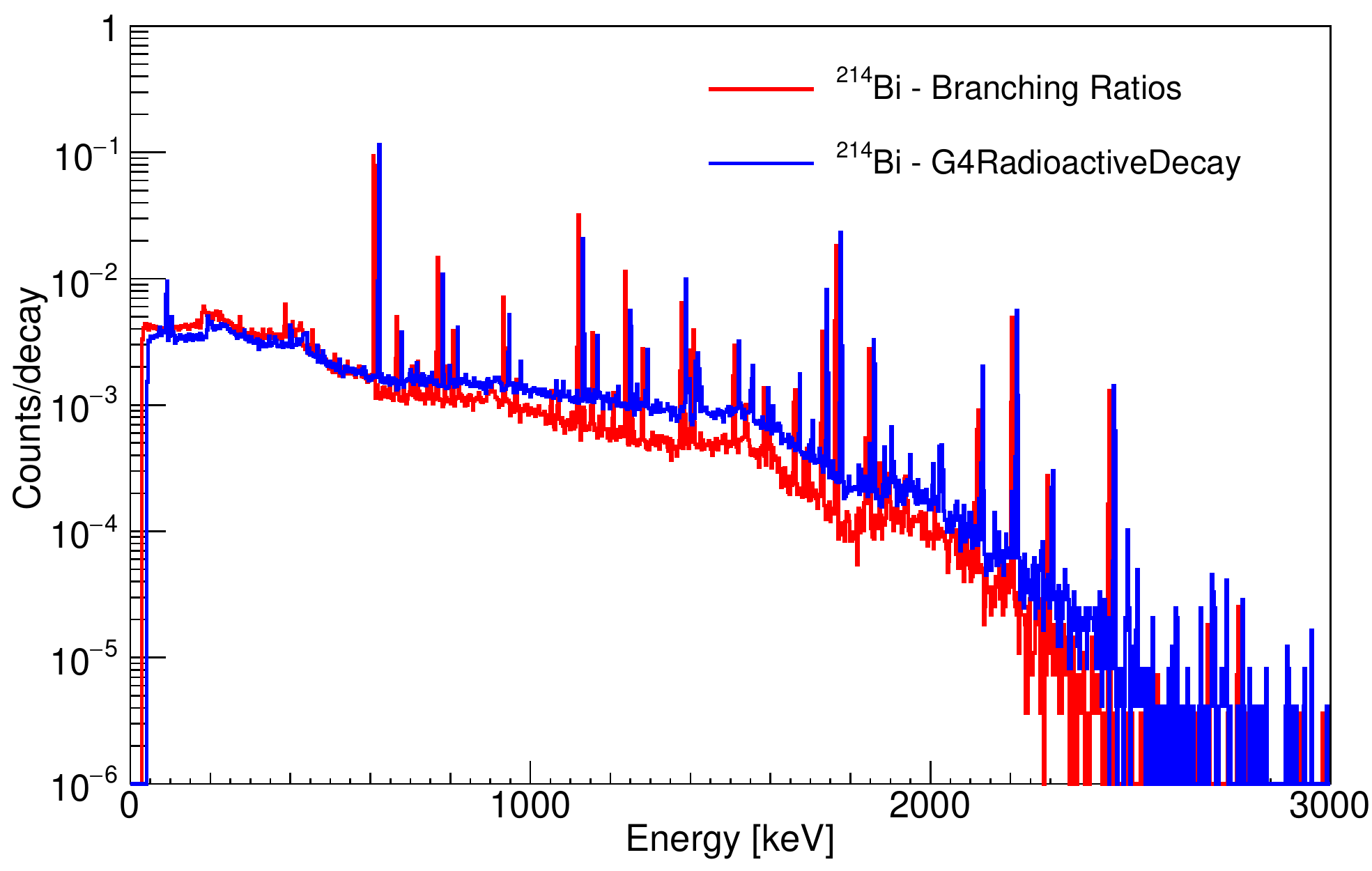}
\caption{A comparison between simulated $^{214}$Bi spectra for a sample in the well of Lumpsey. The red spectrum is produced by a simulation which uses simple \gray\ branching ratios and the blue spectrum is produced using the \geant\ radioactive decay libraries which includes the summing effects described in the main text. For ease of comparison, the blue spectrum is offset by \SI{1}{\keV}. The resolution observed in these spectra is as measured for Lumpsey in Figure~\ref{fig:DetectorResn}.
\label{fig:Bi214CCF}}
\end{figure}

The \geant\ package includes detailed radioactive decay libraries which include information about both \gray\ energies and the half-lives of the energy levels from which they are emitted. To determine the appropriate true coincidence summing correction factors (CCFs), we run a simulation using the full G4RadioactiveDecay libraries limited to a single isotope decay in a U/Th chain. In parallel, we run a second simulation which uses simple \gray\ branching ratios in the same isotopic decay. This second simulation is then scaled to match the average number of \grays\ emitted in the decay in question. A simple fit may then be applied to each peak of interest in the decay simulated by both methods. The ratio of counts calculated using the NNDC~\cite{PRITYCHENKO2011213} values compared to the counts calculated using the G4RadioactiveDecay libraries defines the CCF. \textcolor{black}{The CCF can then be used multiplicatively to correct the measured specific activity of its corresponding \gray\ line}. Figure~\ref{fig:Bi214CCF} shows a comparison between the two libraries for the decay of $^{214}$Bi. The correction factors for $^{214}$Bi have perhaps the greatest impact on detector characterisation as this decay includes several cascade de-excitations which include energy levels with short half-lives. Table~\ref{tab:LumpseyCCF} details the calculated CCFs for several of the \gray\ energies associated with the decay of $^{214}$Bi. It is assumed that any differences between \gray\ intensities in the NNDC and G4RadioactiveDecay libraries are minimal.

\begin{figure}
\centering
  \includegraphics[width=1.0\linewidth]{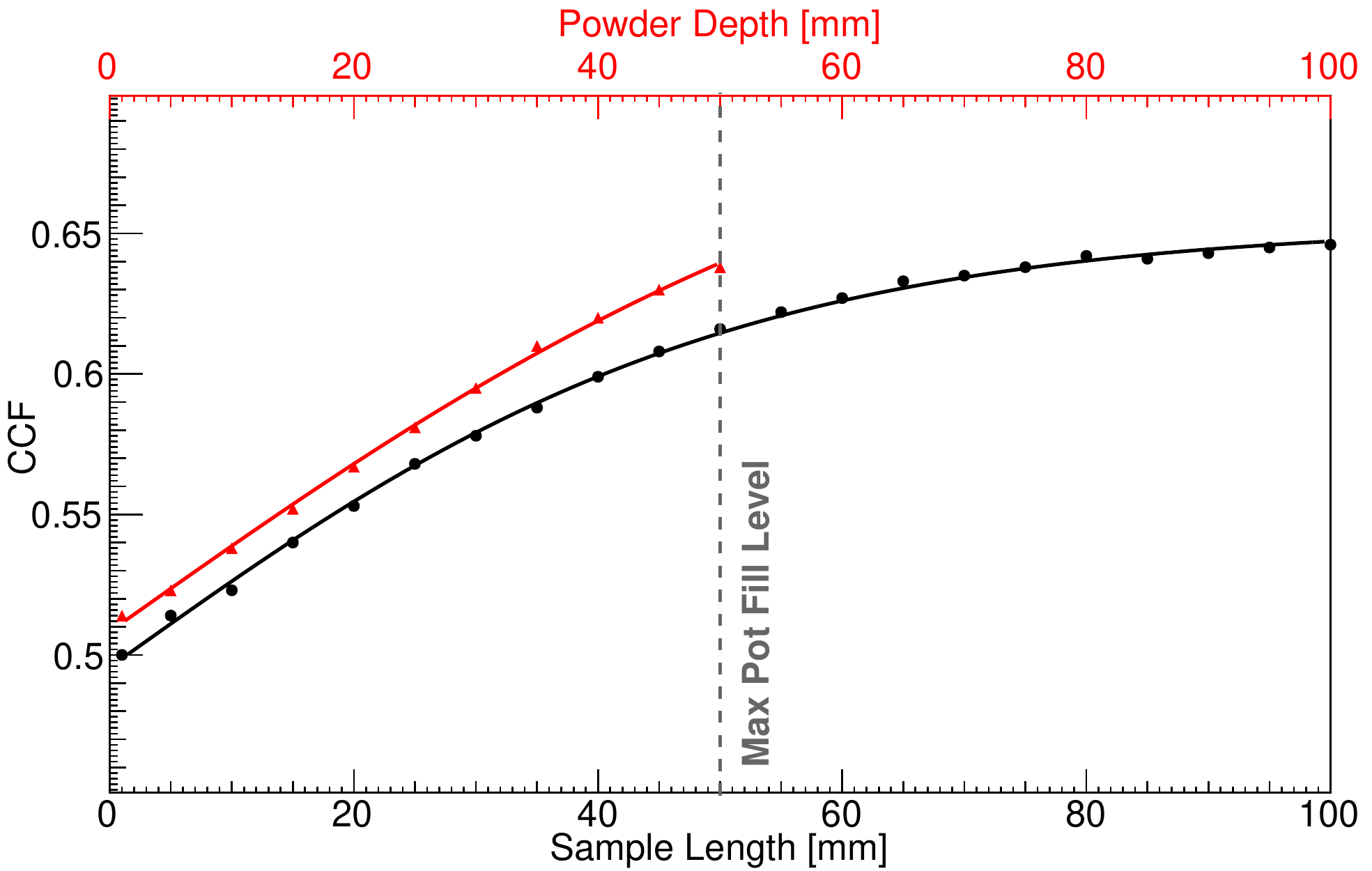}
\caption{Variation in CCF for the \SI{609.3}{\keV} $^{214}$Bi line for a sample placed in the well of Lumpsey as a function of powder depth (red triangle) or PTFE sample length (black circle). The standard Lumpsey well pot has a maximum fill level of \SI{50}{\milli\meter}. The offset in the two curves is due to the differing sample densities and different effective sample geometries (the powder sample is held in a \SI{1}{inch} diameter pot with a wall thickness of \SI{2}{\milli\meter} whereas the PTFE rod has a diameter of \SI{1}{inch}).
\label{fig:CCFDepth}}
\end{figure}

The calculation of CCFs depends strongly on the geometry of the sample measured. Using Lumpsey as an example, the height to which the well is filled determines the effective coverage of the sample by the detector crystal. For a small sample, the coverage is close to 4$\pi$, but for a large sample the coverage is much lower. This means that the probability for multiple \grays\ emitted within a very short time window depositing energy in the crystal decreases with sample size. To illustrate this, Figure \ref{fig:CCFDepth} shows how the CCF for the \SI{609.3}{\keV} line from the decay of $^{214}$Bi varies with sample fill height. Figure \ref{fig:CCFDepth} also shows the CCF as calculated for a PTFE sample of increasing length showing that the CCF continues to vary even for very long samples. This relates to the fact that an increasingly smaller proportion of the sample is within the well.

\begin{table}
  \centering
  \caption{CCFs that are applied to lines in $^{214}$Bi for the \SI{1}{inch} well pot sample on Lumpsey and Chaloner. More substantial correction factors are required in Lumpsey due to the increased solid angle posed by the detector crystal.}
  {\footnotesize
    \begin{tabular}{|C{1.1cm}|C{1.1cm}|C{1.1cm}|}
    \hline
    {{Energy}} & {{CCF}} & {{CCF}}  \\
        {{(keV)}} & {Lumpsey} & {Chaloner}\\
    \hline
    609.3 & 0.64 & 0.87\\
    768.4 & 0.56 & 0.83\\
    934.1 & 0.56 & 0.89\\
    1120.3 & 0.57 & 0.82\\
    1764.5 & 1.02 & 1.01\\
    2204.2 & 1.03 & 1.02\\
    \hline
    \end{tabular}%
    }
  \label{tab:LumpseyCCF}%
\end{table}%

\subsection{Calculating Sample Activity}
\begin{figure}
\centering
\subcaptionbox*{}{%
  \includegraphics[width=1.0\linewidth]{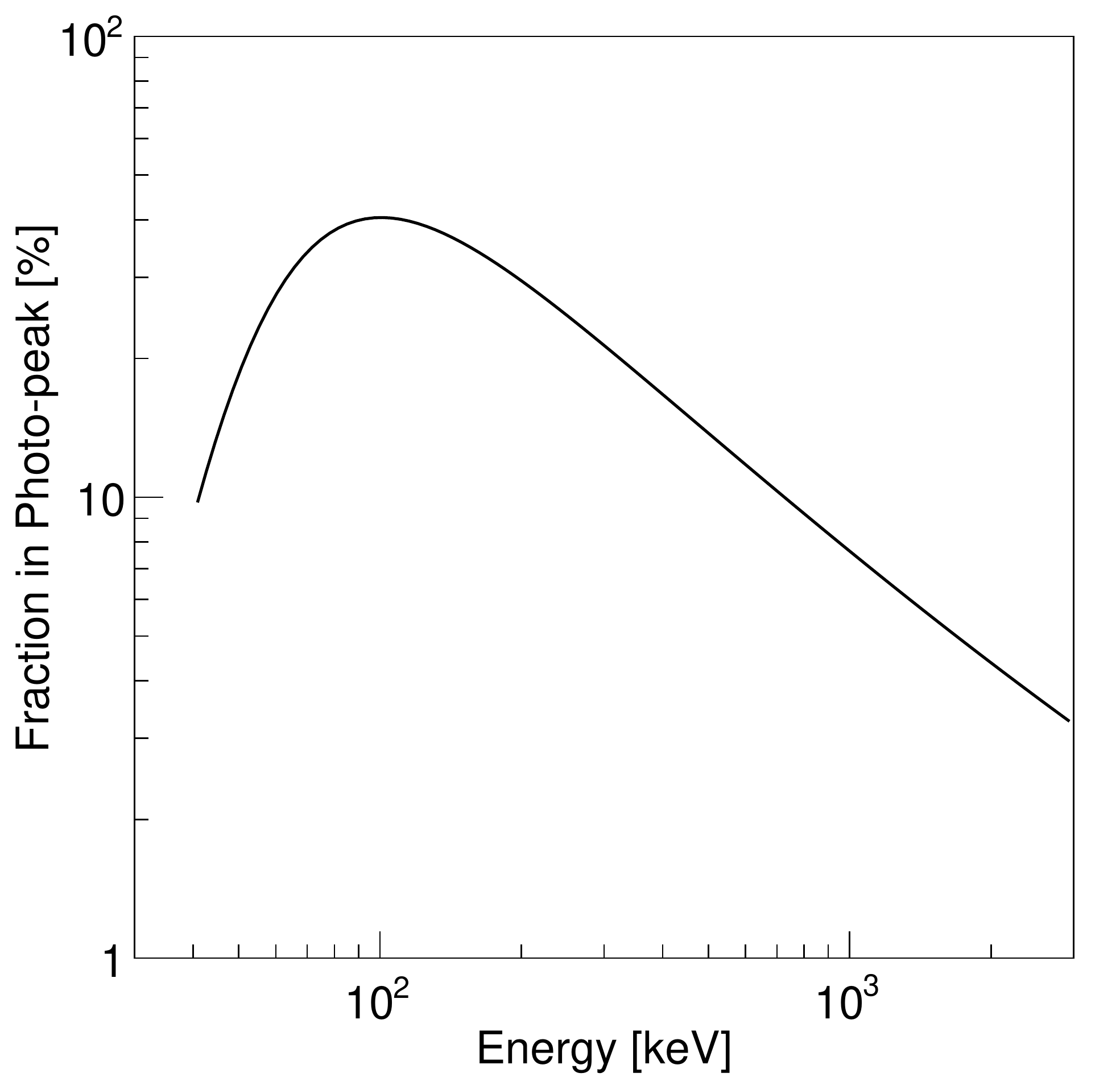}
  \llap{\raisebox{5.cm}{\includegraphics[height=3cm]{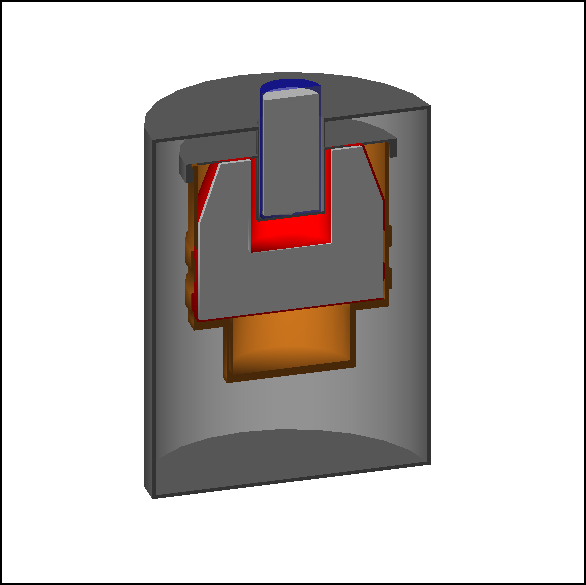}}}   
  }
\subcaptionbox*{}{%
  \includegraphics[width=1.0\linewidth]{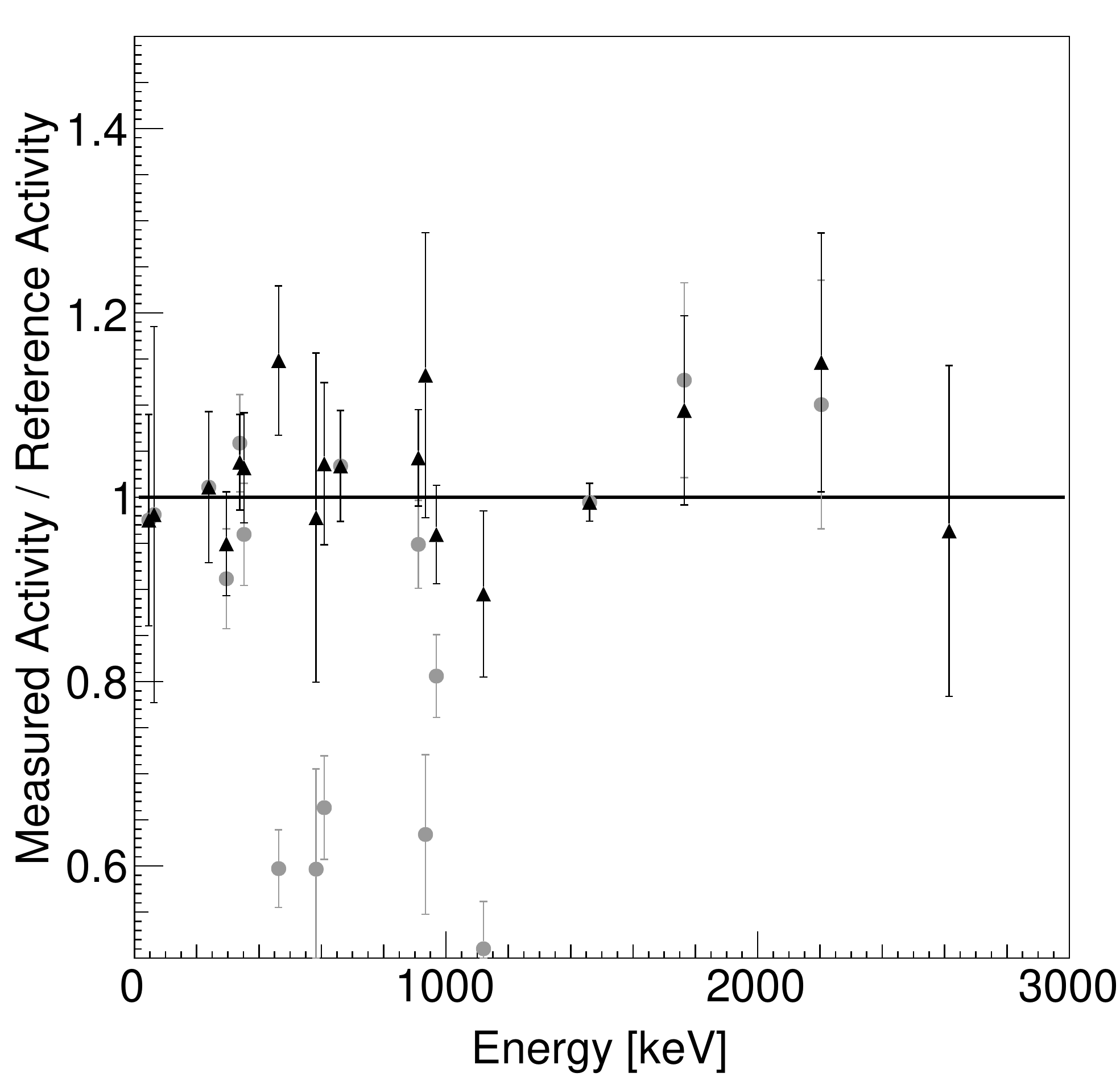}
  }  
\caption{(top) Simulated full photopeak efficiency for a \SI{1}{inch} pot filled to a level of \SI{50}{\milli\meter} and placed in the well of the Lumpsey detector. The simulated geometry is shown in the inset. (bottom) Fractional comparison between measured and reference activity for the IAEA385 sample. The error bars shown are a combination of the uncertainties given on the reference activities and those on the measured activities. All measured activities fall within \SI{15}{\percent} of the reference activities and all but two are statistically within 1$\sigma$ of unity. Grey circles are the original measured values while black triangles are after applying CCFs.
\label{fig:EffWellpotLumpsey}}
\end{figure}

\begin{figure}
\centering
\subcaptionbox*{}{%
  \includegraphics[width=1.0\linewidth]{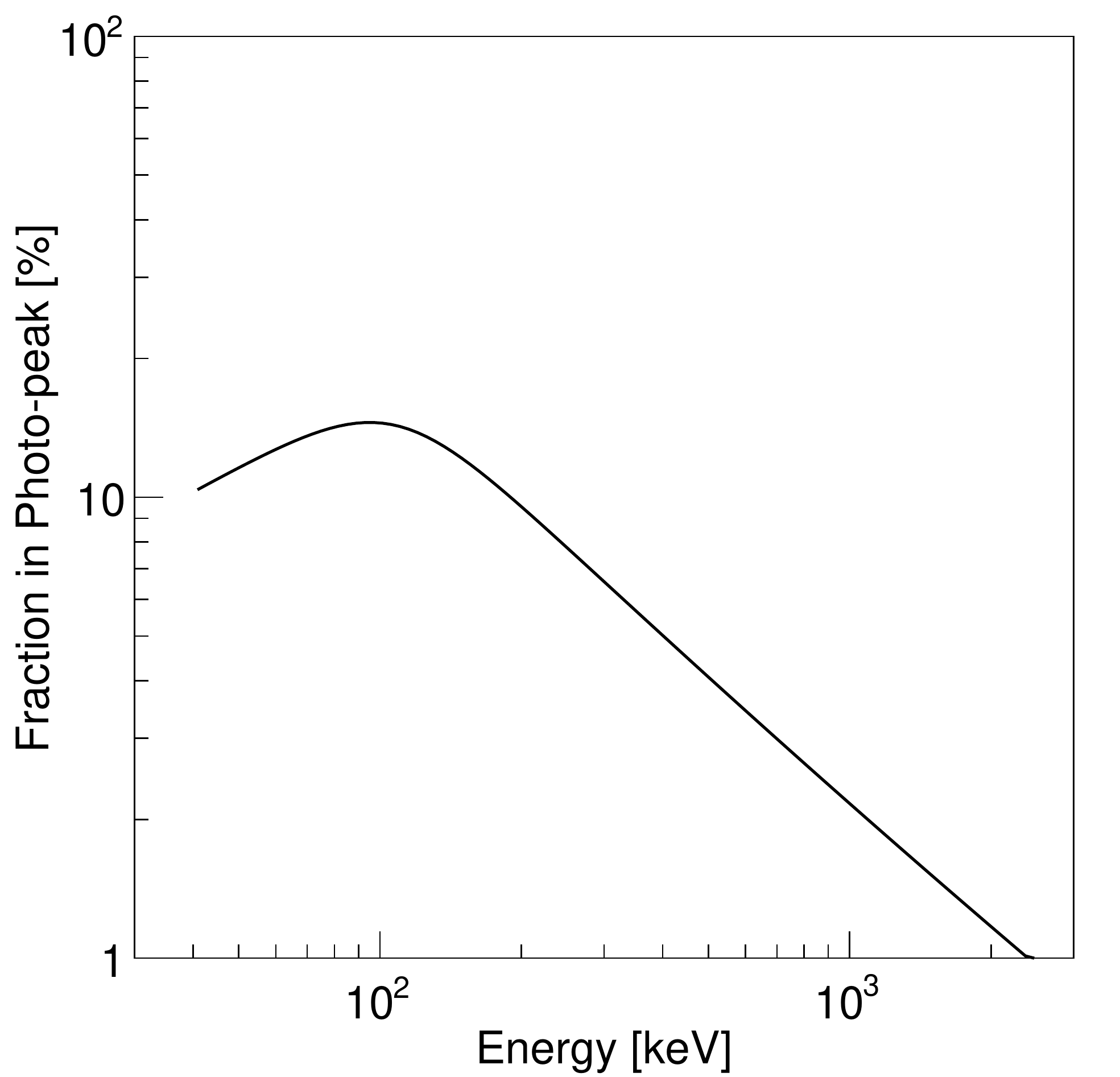}
  \llap{\raisebox{5.cm}{\includegraphics[height=3cm]{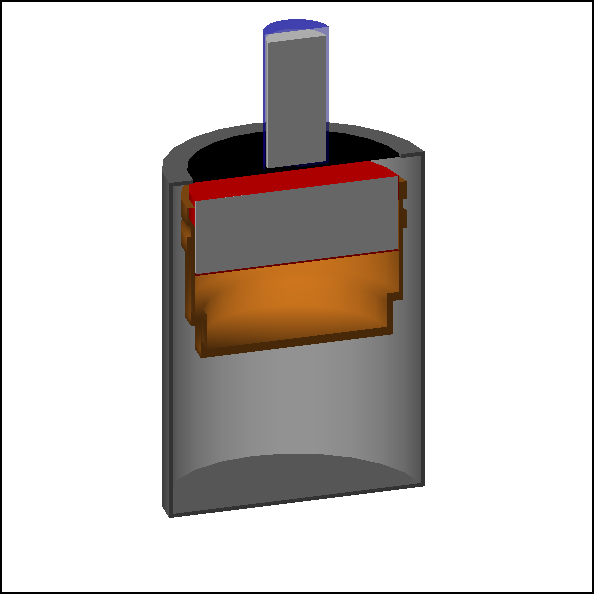}}}   
  }
\subcaptionbox*{}{%
  \includegraphics[width=1.0\linewidth]{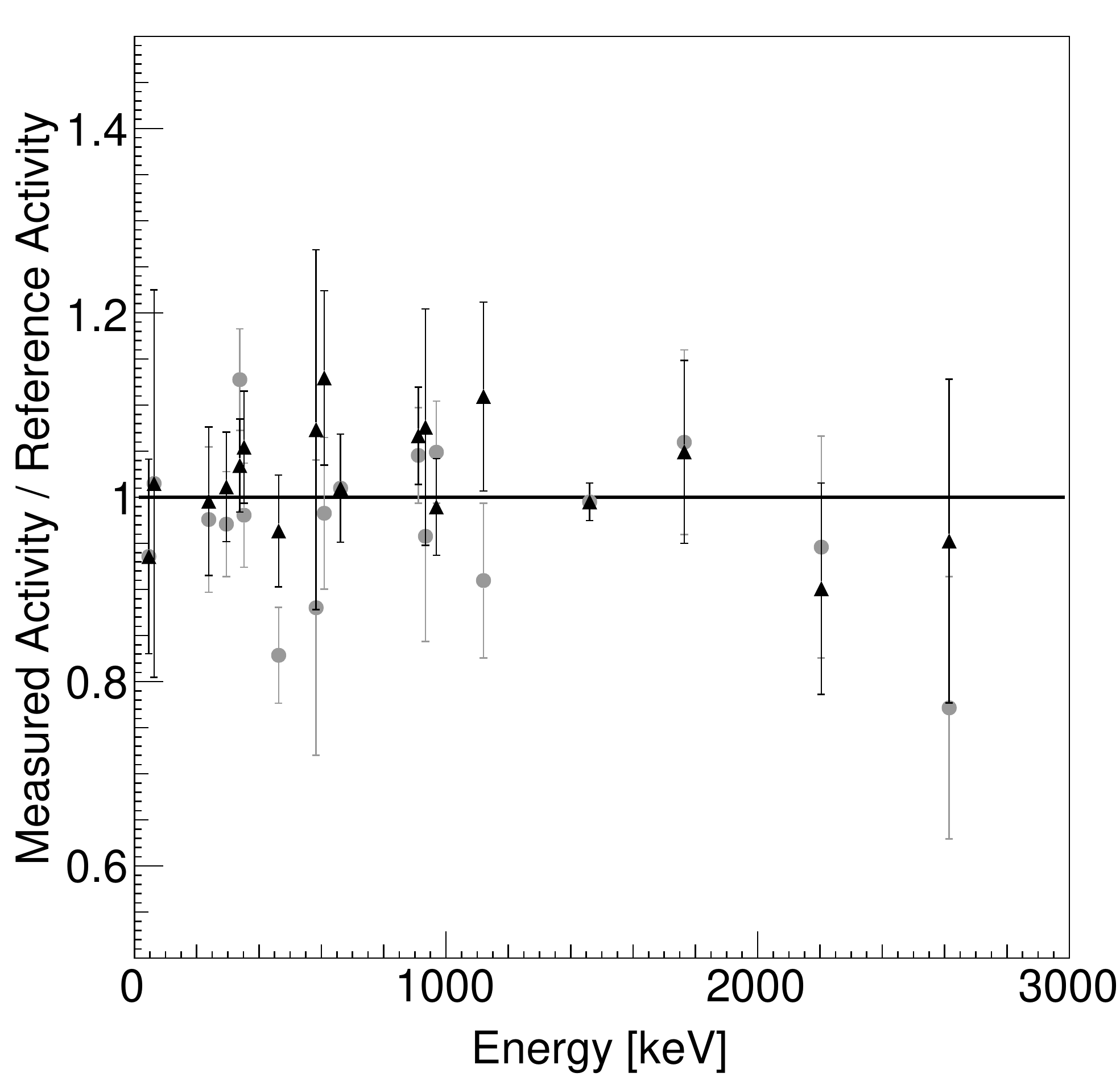}
  }  
\caption{(top) Simulated full photopeak efficiency for a \SI{1}{inch} pot filled to a level of \SI{50}{\milli\meter} and placed at the face of the Chaloner detector. The simulated geometry is shown in the inset. (bottom) Fractional comparison between measured and reference activity for the IAEA385 sample. The error bars shown are a combination of the uncertainties given on the reference activities and those on the measured activities. All measured activities fall within \SI{15}{\percent} of the reference activities and all but three are statistically within 1$\sigma$ of unity. Grey circles are the original measured values while black triangles are after applying CCFs.
\label{fig:EffWellpotChaloner}}
\end{figure}

With CCFs calculated for each of the decays in the IAEA385 sample, it is possible to determine if the calculated efficiency and, hence, the defined detector and sample geometries, give calculated contamination values consistent with those from the IAEA certification. Table \ref{tab:LumpseyIAEA} shows the comparison between those and the values measured using both Lumpsey and Chaloner with agreement seen between all values. Figures~\ref{fig:EffWellpotLumpsey} and~\ref{fig:EffWellpotChaloner} show the ratio of measured to reference activity for each line of interest in Chaloner and Lumpsey, respectively. The errors shown in these figures are purely statistical and, even so, most measurements are compatible with the reference value. The same measurements have been performed with a variety of sample geometries and a similar level of agreement has been observed in all. The certified value for $^{137}$Cs (\SI{33.0 \pm 0.5}{\becquerel\per\kilogram}) has been modified to allow for the time that has passed between certification (01/01/1996) and these measurements.

\begin{table}
  \centering
  \caption{Specific activities for each isotope measured using an identical sample within the well cavity of Lumpsey and on the front face of Chaloner. Not only is there agreement between the two measurement, there is also agreement with the certified values for IAEA385.  }
   {\footnotesize
    \begin{tabular}{|C{1.1cm}|C{1.1cm}|C{1.1cm}|C{1.1cm}|}
    \hline
    	 {{Isotope}} & {Certified value} {(\SI{}{\becquerel\per\kilogram})} 
	 & {Lumpsey} {(\SI{}{\becquerel\per\kilogram})} & {Chaloner} {(\SI{}{\becquerel\per\kilogram})} \\
    \hline
\rule{0pt}{2.4ex}$^{40}$K & $608 \pm 6$ & $611 \pm 6$ & $611 \pm 11$ \\
$^{137}$Cs & $21.9 \pm 0.3$ & $21.4 \pm 0.3$ & $21.2 \pm 1.2$ \\
$^{208}$Tl & $11.3 \pm 0.3$ & $11.9 \pm 0.3$ & $11.6 \pm 2.1$ \\
$^{210}$Pb & $34.6 \pm 1.4$ & $33.2 \pm 0.9$ & $35.5 \pm 3.9$ \\
$^{212}$Bi & $34.9 \pm 1.4$ & $34.5 \pm 1.9$ & $34.2 \pm 1.2$ \\
$^{212}$Pb & $37.5 \pm 0.4$ & $37.1 \pm 0.3$ & $37.3 \pm 3.0$ \\
$^{214}$Bi & $20.0 \pm 0.7$ & $21.7 \pm 0.7$ & $19.6 \pm 1.6$ \\
$^{214}$Pb & $21.4 \pm 0.4$ & $22.3 \pm 0.4$ & $21.6 \pm 1.2$ \\
$^{228}$Ac & $32.6 \pm 1.0$ & $32.1 \pm 0.8$ & $31.5 \pm 1.4$ \\
$^{234}$Th & $28.2 \pm 0.9$ & $29.1 \pm 0.7$ & $28.7 \pm 5.9$ \\
\hline
    \end{tabular}%
    }
  \label{tab:LumpseyIAEA}%
\end{table}%

\subsection{Comparison with Other Detectors}

As part of the LZ materials screening and selection programme, both Chaloner and Lunehead have been cross-calibrated with detectors at the Black Hills Underground Campus (BHUC) of the Sandford Underground Research Facility (SURF) and detectors at the University of Alabama. This cross-calibration was performed using a sample of Rhyolite of known specific activity that was screened using each detector in turn. Both Chaloner and Lunehead measured contaminations within the error of the known activity. Overall the cross-calibration puts all measurements for all isotopes within \SI{12}{\percent} across 9 detectors.

\subsection{Detector Relative Efficiency}

With the three detectors fully characterised, it is possible, through simulation, to recreate the method by which their relative efficiencies are measured. In order to do this, a simple model of a \SI{3x3}{inch} NaI detector is constructed in \geant\ to run a simulation with a \cosz\ source held at \SI{25}{\cm} from the detector face. The same simulation is then performed for the three BUGS detectors with the ratio of these results and that from the NaI detector giving the measure of relative efficiency. Both Chaloner and Lumpsey give identical results to that given in their characterisation documentation from Canberra. The dead-layer profile of Lunehead (where a large proportion of the detector crystal must be inactive) leads to a simulated relative efficiency of \SI{71}{\percent} as compared to the \SI{92}{\percent} that a crystal of its size would originally possess. The cause of this difference may be assessed using the detector response to \kfz\ decays. The vast majority of \kfz\ \gray s originate in the cavern rock surrounding the laboratory. In both the pre- and post-refurbishment spectra a similar shielding configuration was used and the rate of \kfz\ in the background spectrum is identical with $\SI{210 \pm 21}{\micro\becquerel}$ pre-refurbishment and $\SI{210 \pm 13}{\micro\becquerel}$ post-refurbishment (see Figure~\ref{fig:lunhead_improv}). This agreement is not consistent with the rate loss that would be expected if the refurbishment had caused a significant decrease in the active volume of the detector crystal. The reduction in relative efficiency is, therefore, thought to be due to Lunehead being previously stored warm for extended periods allowing the diffusion of lithium atoms into the germanium crystal~\cite{boson}.

\section{Detector Sensitivity to Realistic Samples}

\subsection{Detector Backgrounds}

\begin{figure}
\centering
  \includegraphics[width=1.0\linewidth]{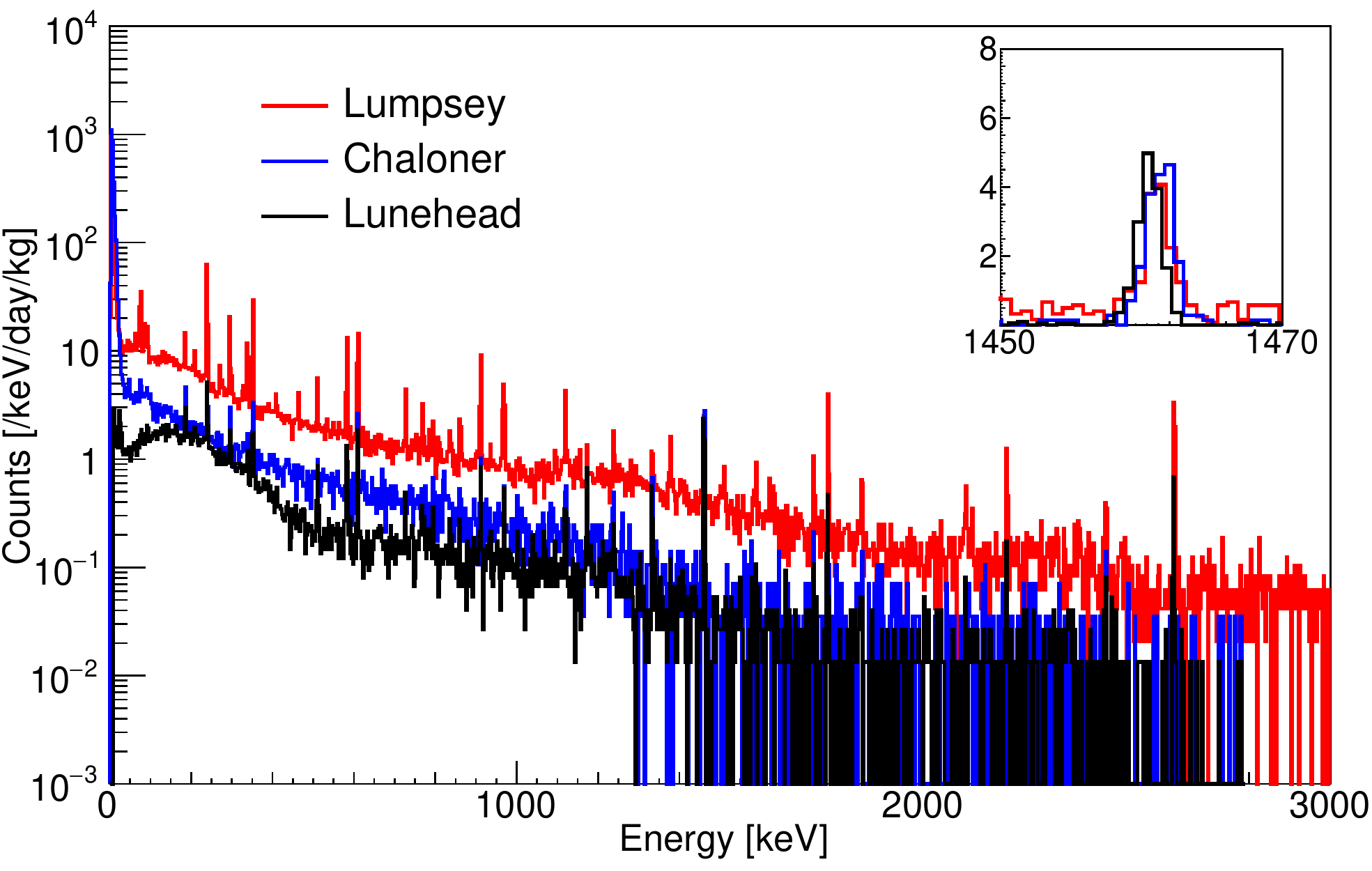}
\caption{Measured background spectra for the three BUGS detectors. The increased background seen in Lumpsey is expected. This is due to the fact that some intrinsically higher background components must be used in the construction of the detector in order to maximise the quality of the detector resolution. The inset plot \textcolor{black}{(which uses the same axis values as the main plot)} zooms into the \kfz\ line at \SI{1460.8}{\kilo\electronvolt} which clearly shows that the mass corrected background rate is independent of detector geometry and composition. 
\label{fig:DetectorBkg}}
\end{figure}

The three Boulby detectors exhibit backgrounds that make them suitable for the screening of radio-pure materials for low-background DM and \dbd\ experiments. Figure~\ref{fig:DetectorBkg} shows the backgrounds from each of the detectors in terms of events per day per keV scaled to the detector crystal mass. A study of these backgrounds in combination with the simulated detector response can be used to determine a minimum detectable activity (MDA) for each detector. With geometries being equal, the calculated MDA depends on the density of the sample. For high density samples, better sensitivity is achieved for higher energy \grays\ where the increase in mass dominates over the efficiency loss due to self-absorption (where a \gray\ deposits some or all of its energy within the sample being measured). For lower density samples on a BEGe-type detector (Chaloner and Lumpsey well) the increase in efficiency to low-energy \grays\ dominates the reduction in mass. An interesting feature of these spectra is the \kfz\ rate as judged by the \gray\ at \SI{1460.8}{\kilo\electronvolt}. This shows that the rate per unit mass is relatively independent of detector. The three shields are almost identical so this suggests that the rate we see is from a source external to the shield.

\subsection{Calculating Sensitivity}
\label{sec:mds}

For comparison, we look at two materials traditionally used in the construction of low-background experiments: PTFE and copper. In both cases, we simulate a large disk sample and a smaller sample of \SI{1}{inch} diameter. In addition to these samples, and for Lunehead only, we can calculate the MDA for a powder sample in a \SI{1}{\litre} Marinelli beaker. The crystal geometries of Chaloner and Lumpsey mean that little benefit is gained from surrounding the crystal with sample material.

\begin{figure}
\centering
  \includegraphics[width=1.0\linewidth]{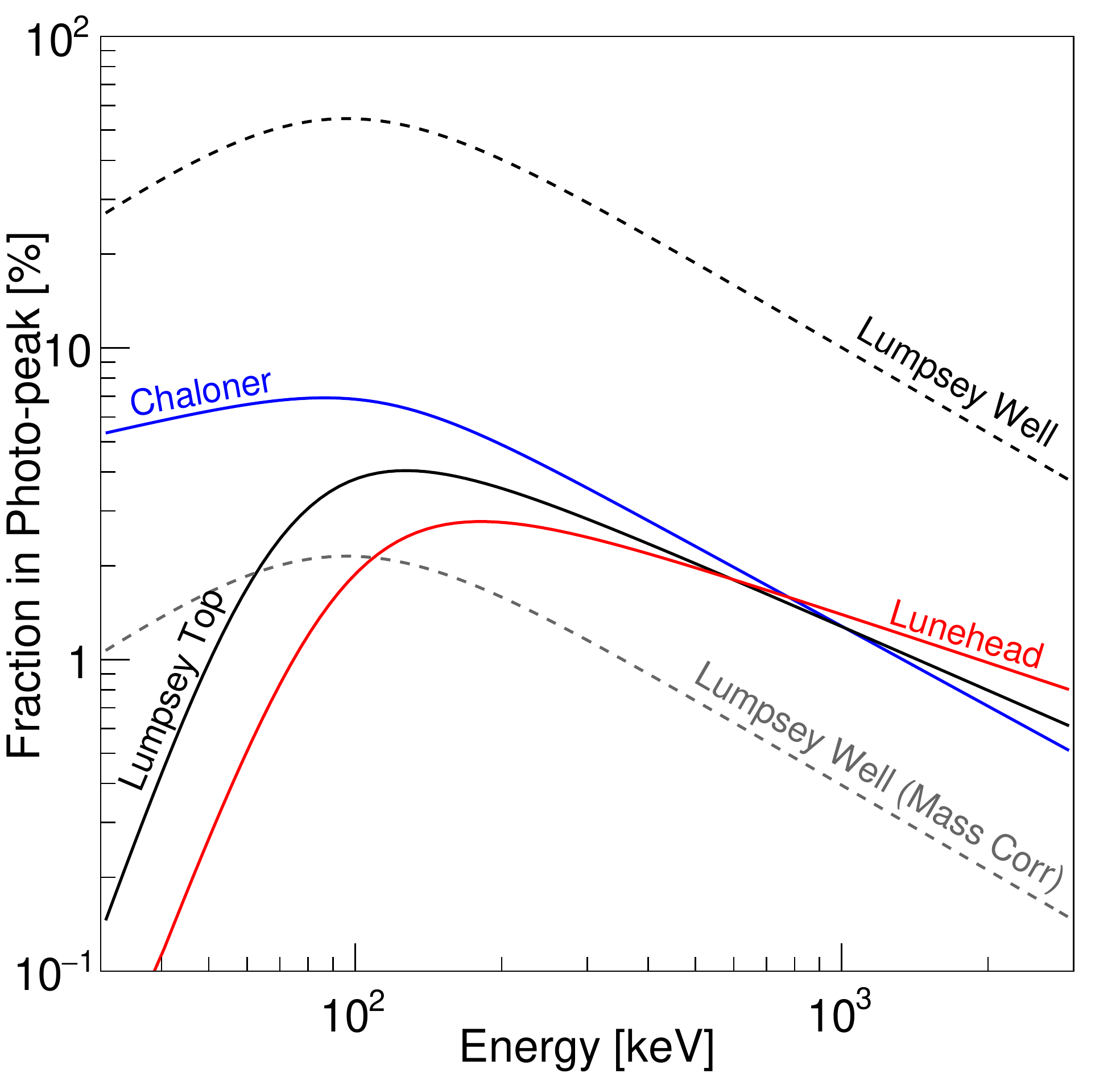}
\caption{Calculated efficiencies for the representative PTFE geometries. The \SI{850}{\g} sample efficiency is shown for Chaloner (blue), Lunehead (red) and Lumpsey (solid black). The efficiency in the Lumpsey well for the \SI{34}{\g} sample is shown with a dashed black line. Additionally, the dashed grey line represents the Lumpsey well efficiency when scaled to allow for the smaller sample mass as described in the text. This shows that, although the well has a much higher efficiency for samples, this is counteracted by the reduction in mass.
\label{fig:PTFEEff}}
\end{figure}

\begin{figure}
\centering
  \includegraphics[width=1.0\linewidth]{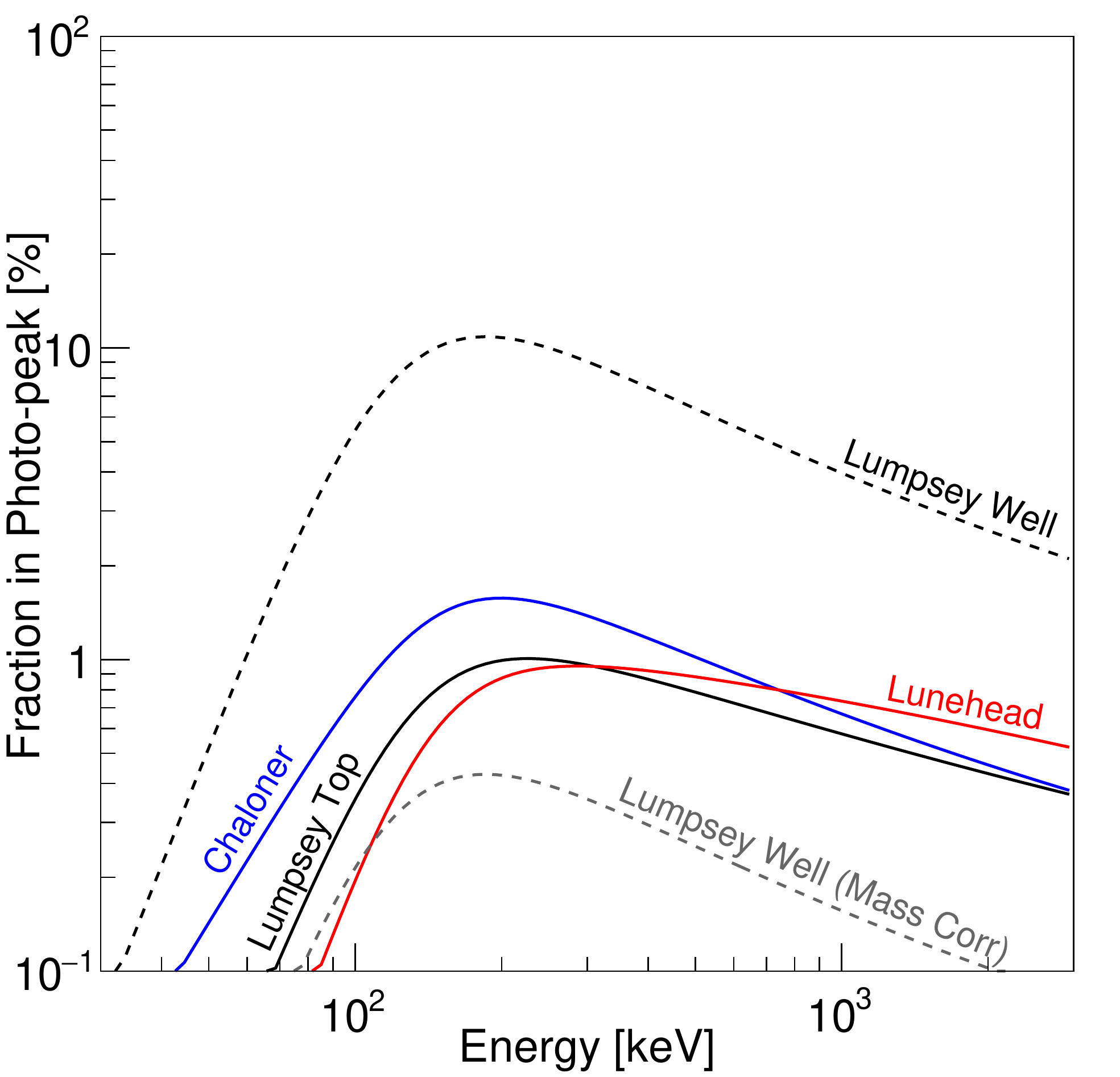}
\caption{Calculated efficiencies for the representative copper geometries. The \SI{3.5}{\kg} sample efficiency is shown for Chaloner (blue), Lunehead (red) and Lumpsey (solid black). The efficiency in the Lumpsey well for the \SI{240}{\g} sample is shown with a dashed black line. Additionally, the dashed grey line represents the Lumpsey well efficiency when scaled to allow for the smaller sample mass as described in the text. When compared with the efficiency for PTFE samples in Figure~\ref{fig:PTFEEff}, it is clear that the low-energy efficiency is severely suppressed and falls off in the BEGe-type detectors at almost the same rate as the coaxial detectors. At the same time, the higher energy efficiency only suffers a relatively modest decrease when compared with PTFE.
\label{fig:CuEff}}
\end{figure}

Figures~\ref{fig:PTFEEff} and~\ref{fig:CuEff} show the calculated efficiencies for the PTFE and copper samples, respectively. In order to give a farer comparison between the efficiency of a sample placed in the well of Lumpsey and that of a sample placed on the face of all other detectors, a scaling factor determined by the mass difference is applied to the efficiency curve of the well detector. This shows that, although there is a much higher efficiency for these samples, the lower mass seriously impacts the overall sensitivity. For example, with equal background rates, a large sample on Chaloner will give a lower MDA than a small sample in the well of Lumpsey regardless of energy. Of course, in the case where only a small sample is available, Lumpsey regains the advantage.

Tables~\ref{tab:DetectorMDAPTFE} and~\ref{tab:DetectorMDACu} detail the calculated \SI{90}{\percent} confidence level (C.L.) MDAs for all sample types based on a screening duration of 14 days. This duration is chosen as a balance between reaching the ultimate sensitivity achievable and maximising sample throughput. It is clear that the BEGe-type detectors (Chaloner and Lumpsey well) are more sensitive to both \utTeE\ and \pbtoz\ in the lower density PTFE samples. In the case of \utTeE, this is helped by the fact that the 63.3~keV \gray, which has a higher relative intensity that the \SI{1001}{\keV} line, may be used. For copper, the efficiency at \SI{63.3}{\keV} is so low that a determination of MDA may only be performed using the \SI{1001}{\keV} line. In both cases, we assume that any contamination of \pbtoz\ is found in the bulk of the material screened rather than on the surface. It is useful to note that the line used to set an MDA for \thtTtL, \SI{583}{\keV}, comes from the decay of \tltze. The levels of this isotope in materials is of particular interest to the \dbd\ community as the other prominent \tltze\ decay \gray\ is at \SI{2615}{\keV}. Many potential \dbd\ channels have $Q_{\beta\beta}$ of \SI{\sim2}{\MeV}~\cite{GiuliDBD} which means any potential signal from \dbd\ may be dominated by Compton scatters from the \SI{2615}{\keV} \gray\ if materials without the required purity are used in experimental construction.

Table~\ref{tab:DetectorMDAPowder} details the calculated \SI{90}{\percent} C.L. MDAs for the \SI{1}{\litre} Marinelli beaker on Lunehead calculated for a screening duration of 14 days. We use the same material and density as used in the IAEA385 simulations to represent the powdered sample. This gives a sample mass of \SI{1.2}{\kilo\gram}. The results show that the improved efficiency of a Marinelli beaker balances the mass lost as compared to the copper sample with the resultant MDAs coming out similar in both cases.

To summarise the sensitivity of the three BUGS detectors, we take the best possible combination of MDAs from the PTFE and copper sample to give the following:
\begin{itemize}
\item \utTeE: \SI{3.6}{\milli\becquerel\per\kilo\gram} (\SI{290e-12}{g/g})
\item \utTeL: \SI{0.9}{\milli\becquerel\per\kilo\gram} (\SI{70e-12}{g/g})
\item \pbtoz: \SI{6.6}{\milli\becquerel\per\kilo\gram} (\SI{530e-12}{g/g})
\item \utTF: \SI{0.9}{\milli\becquerel\per\kilo\gram} (\SI{1500e-12}{g/g})
\item \thtTtE: \SI{1.1}{\milli\becquerel\per\kilo\gram} (\SI{270e-12}{g/g})
\item \thtTtL: \SI{0.3}{\milli\becquerel\per\kilo\gram} (\SI{70e-12}{g/g})
\item \kfz: \SI{8.6}{\milli\becquerel\per\kilo\gram} (\SI{270e-9}{g/g})
\end{itemize}
where we assume in the conversion from \SI{}{\milli\becquerel\per\kilo\gram} to \SI{}{g/g} of the parent isotope in the decay chain that later chain isotopes are in secular equilibrium with their respective parents (\utTe\ and \thtTt).

\begin{table}
  \centering
  \caption{Calculated \SI{90}{\percent} C.L. MDAs (\SI{}{\milli\becquerel\per\kilogram}) for a typical PTFE sample on each of the BUGS detectors. MDAs are calculated for the two possible configurations of Lumpsey --- with a large sample on the face [1] and with a smaller sample in the well [2]. The energies used in the MDA calculation are also highlighted \textcolor{black}{with the symbols found next to MDA values corresponding to those next to the \gray\ energies.}}
    {\footnotesize
    \begin{tabular}{C{1.1cm}|C{1.1cm}|C{1.1cm}|C{1.1cm}|C{1.1cm}|}
    \cline{2-5}
    \rule{0pt}{2.4ex}& \multicolumn{4}{ c| }{{all \SI{90}{\percent} C.L. MDAs in \SI{}{\milli\becquerel\per\kilogram}}} \\ 
    \hline
     \multicolumn{1}{ |c| }{{Detector} }
     & \rule{0pt}{2.4ex}{\utTeE\ 63.3~keV$^{\star}$ 1001~keV$^{\dagger}$} 
     & {\utTeL\ 351.9~keV$^{\star}$ 609.3~keV$^{\dagger}$} 
     & {{\pbtoz\ 46.5~keV}} 
     &  {{\utTF\ 143.8~keV}}\\
    \hline
       \multicolumn{1}{ |c| }{ Chaloner} & 3.6$^{\star}$ & 1.7$^{\star}$ & 6.6 & 0.9\\
       \multicolumn{1}{ |c| }{ Lunehead} & 17.0$^{\dagger}$ & 2.4$^{\dagger}$ & - & 2.7\\ 
       \multicolumn{1}{ |c| }{ Lumpsey [1]} & 56.4$^{\dagger}$ & 7.6$^{\dagger}$ & - & 3.8\\
       \multicolumn{1}{ |c| }{ Lumpsey [2]} & 30.2$^{\star}$ & 21.4$^{\star}$ & 49.7 & 7.7\\
    \hline\hline
         \multicolumn{1}{ |c| }{{Detector} } 
     & \rule{0pt}{2.4ex}{{\thtTtE\ 338.3~keV$^{\star}$ 911.2~keV$^{\dagger}$}} 
     & {{\thtTtL\ 583.2~keV}} 
     & {{\kfz\ 1461~keV}} 
     &  {{\cosz\ 1173~keV}} \\
     \hline
        \multicolumn{1}{ |c| }{ Chaloner} & 2.1$^{\star}$ & 0.5 & 20.2 & 0.7\\
        \multicolumn{1}{ |c| }{ Lunehead} & 3.5$^{\star}$ & 1.2 & 23.4 & 1.0\\
        \multicolumn{1}{ |c| }{ Lumpsey [1]} & 11.9$^{\dagger}$ & 3.3 & 21.1 & 1.1\\ 
        \multicolumn{1}{ |c| }{ Lumpsey [2]} & 34.9$^{\star}$ & 9.6 & 70.9 & 3.8\\
	\hline
    \end{tabular}%
    }
  \label{tab:DetectorMDAPTFE}%
\end{table}%

\begin{table}
  \caption{Calculated \SI{90}{\percent} C.L. MDAs (\SI{}{\milli\becquerel\per\kilogram}) for a typical copper sample on each of the BUGS detectors. MDAs are calculated for the two possible configurations of Lumpsey --- with a large sample on the face [1] and with a smaller sample in the well [2]. Due to the high density of copper compared to PTFE, there is no ambiguity in which line should be used to calculate the MDA for each isotope.}
    {\footnotesize
    \begin{tabular}{C{1.1cm}|C{1.1cm}|C{1.1cm}|C{1.1cm}|C{1.1cm}|}
    \cline{2-5}
    \rule{0pt}{2.4ex}& \multicolumn{4}{ c| }{{all \SI{90}{\percent} C.L. MDAs in \SI{}{\milli\becquerel\per\kilogram}}} \\ 
    \hline
     \multicolumn{1}{ |c| }{{Detector} }
     & \rule{0pt}{2.4ex}{\utTeE\ 1001~keV} 
     & {\utTeL\ 609.3~keV} 
     & {{\pbtoz\ 46.5~keV}} 
     &  {{\utTF\ 143.8~keV}}\\
    \hline
       \multicolumn{1}{ |c| }{Chaloner} & 18.3$^{}$ & 0.9$^{}$ & 88.9 & 1.0 \\
       \multicolumn{1}{ |c| }{Lunehead} & 32.7$^{}$ & 1.2$^{}$ & - & 3.1 \\
       \multicolumn{1}{ |c| }{Lumpsey [1]} & 30.4$^{}$ & 4.4$^{}$ & - & 4.8 \\
       \multicolumn{1}{ |c| }{Lumpsey [2]} & 65.4$^{}$ & 8.6$^{}$ & 950 & 5.6 \\
              \hline\hline
       \multicolumn{1}{ |c| }{{Detector} } 
     & \rule{0pt}{2.4ex}{{\thtTtE\ 911.2~keV}} 
     & {{\thtTtL\ 583.2~keV}} 
     & {{\kfz\ 1461~keV}} 
     &  {{\cosz\ 1173~keV}} \\
     \hline
       \multicolumn{1}{ |c| }{Chaloner} & 1.1$^{}$ & 0.3 & 8.6 & 0.3 \\
       \multicolumn{1}{ |c| }{Lunehead} & 1.7$^{}$ & 0.6 & 9.9 & 0.5 \\
       \multicolumn{1}{ |c| }{Lumpsey [1]} & 6.5$^{}$ & 2.0 & 10.3 & 0.6 \\
       \multicolumn{1}{ |c| }{Lumpsey [2]} & 13.9$^{}$ & 3.8 & 24.2 & 1.3 \\
                     	\hline
    \end{tabular}%
    }
  \label{tab:DetectorMDACu}%
\end{table}%

\begin{table}
  \centering
  \caption{Calculated \SI{90}{\percent} C.L. MDAs (\SI{}{\milli\becquerel\per\kilogram}) for a \SI{1}{\litre} Marinelli beaker filled with IAEA385 powder on the Lunehead detector.}
     { \footnotesize
    \begin{tabular}{C{1.0cm}|C{1.1cm}|C{1.1cm}|C{1.1cm}|C{1.1cm}|}
    \cline{2-5}
    \rule{0pt}{2.4ex}& \multicolumn{4}{ c| }{{all \SI{90}{\percent} C.L. MDAs in \SI{}{\milli\becquerel\per\kilogram}}} \\ 
    \hline
     \multicolumn{1}{ |c| }{{Detector} }
     & \rule{0pt}{2.4ex}{\utTeE\ 1001~keV} 
     & {\utTeL\ 609.3~keV} 
     & {{\pbtoz\ 46.5~keV}} 
     &  {{\utTF\ 143.8~keV}}\\
    \hline
       \multicolumn{1}{ |c| }{Lunehead} & 38.0$^{}$ & 1.2$^{}$ & - & 1.4 \\
              \hline\hline
       \multicolumn{1}{ |c| }{{Detector} } 
     & \rule{0pt}{2.4ex}{{\thtTtE\ 911.2~keV}} 
     & {{\thtTtL\ 583.2~keV}} 
     & {{\kfz\ 1461~keV}} 
     &  {{\cosz\ 1173~keV}} \\
     \hline
       \multicolumn{1}{ |c| }{Lunehead} & 1.9$^{}$ & 0.6 & 12.8 & 0.5 \\
                     	\hline
    \end{tabular}%
    }
  \label{tab:DetectorMDAPowder}%
\end{table}%

\section{Benefits of Broad Energy Range for Real Samples}
\subsection{Significant Disequilibrium at \pbtoz}
\begin{figure}
\centering
  \includegraphics[width=1.0\linewidth]{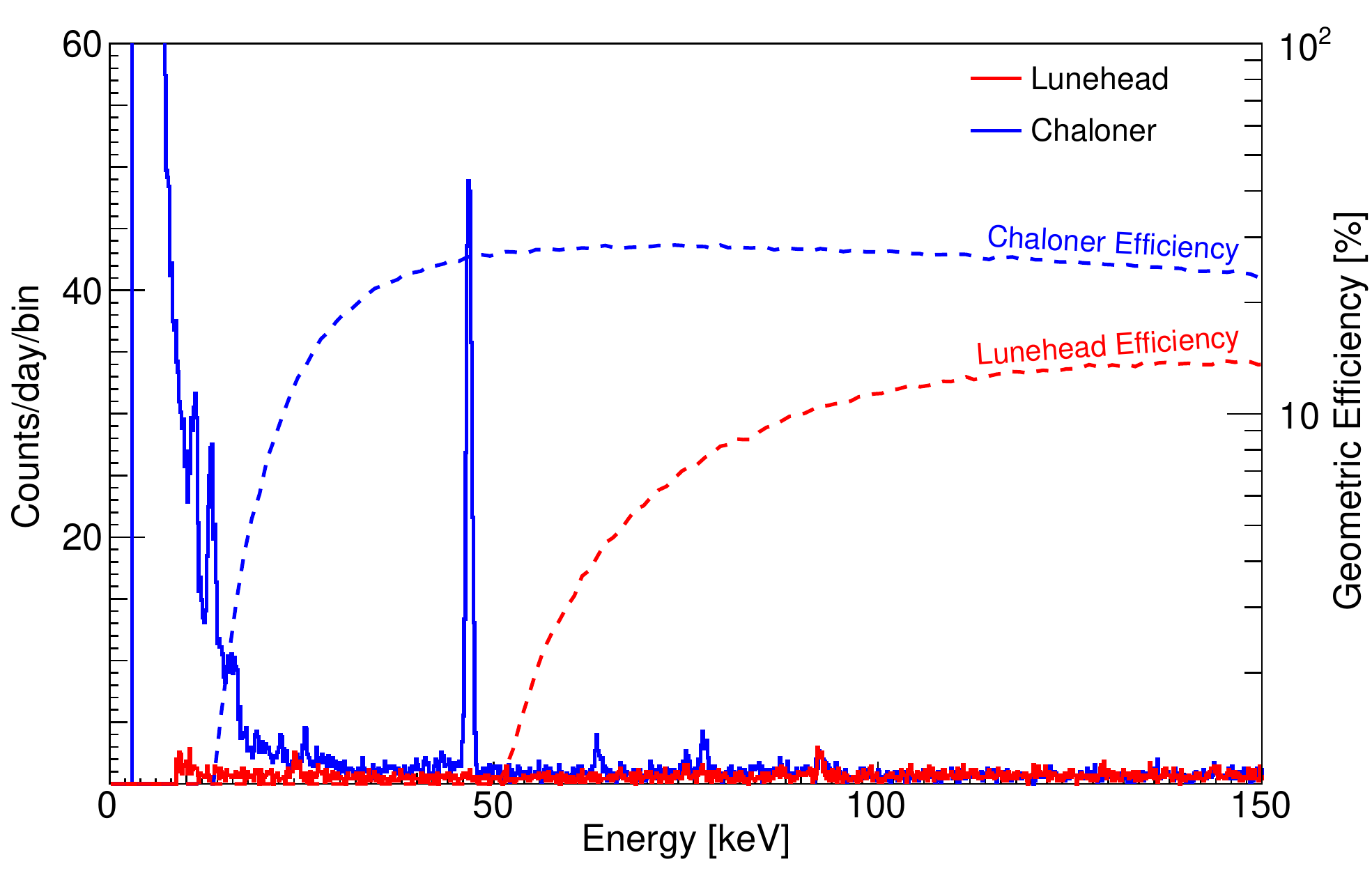}%
  \caption{Comparison between normalised spectra from resistor samples run on Lunehead (red) and Chaloner (blue). It is clear that there is a significant peak \pbtoz\ seen at \SI{46}{\kilo\electronvolt} in Chaloner that is not reproduced in Lunehead. For comparison, the full photopeak efficiency for the sample on each detector is plotted using dashed lines in the same colour as the spectra. It is clear that Chaloner maintains high efficiency at the \pbtoz\ \gray\ energy whereas the efficiency has fallen off precipitously in Lunehead.  The measured specific activity for \pbtoz, shown in Table~\ref{tab:resistors}, is significantly out of equilibrium with the measured value for \utTeL.
\label{fig:resistors}}
\end{figure}

Over the time that BUGS has been routinely screening samples for low-background experiments, the characterisation of several materials has benefitted significantly from the broad energy range of detectors at Boulby. Perhaps the most notable to this end was the screening of a large quantity of resistors provided by the LZ experiment. One sample was specifically split into two batches and screened in parallel using both Chaloner and Lunehead. A comparison between the low-energy bins of both spectra is shown in Figure~\ref{fig:resistors} and specific activity results are shown in Table~\ref{tab:resistors}. Excellent agreement is seen between \utTeL\ in both Chaloner and Lunehead and between the measurements for \utTeE\ and \utTeL\ in Chaloner. However, there is a large disagreement between these values and that measured for \pbtoz\ using Chaloner.  It is only thanks to the sensitivity of the Chaloner BEGe detector to these low-energy \grays\ that we are able to report this effect. 

\begin{table}
  \centering
  \caption{Calculated contaminations for a sample of resistors screened using both Chaloner and Lunehead. A high level of agreement is seen across the \utTe\ chain with the exception of \pbtoz\ which has a measured specific activity 58$\times$ higher than those isotopes above it in the chain (determined using the combined Lunehead/Chaloner measurement of \utTeL). A measurement by Lunehead alone would not observe this.}
    {\footnotesize
    \begin{tabular}{C{1.1cm}|C{1.1cm}|C{1.1cm}|C{1.1cm}|C{1.1cm}|}
    \cline{2-5}
    \rule{0pt}{2.4ex}& \multicolumn{4}{ c| }{{all values in \SI{}{\micro\becquerel\per resistor}}} \\ 
    \hline
     \multicolumn{1}{ |c| }{{Detector} }
     & \rule{0pt}{2.4ex}{\utTeE\ } 
     & \rule{0pt}{2.4ex}{\utTeL\ } 
     & \rule{0pt}{2.4ex}{\pbtoz\ } 
     & \rule{0pt}{2.4ex}{\utTF\ }\\
    \hline
       \multicolumn{1}{ |c| }{ Chaloner} & 5.8$\pm$1.7 & 3.7$\pm$0.9 & 267$\pm$9 & $<$0.3 \\
       \multicolumn{1}{ |c| }{ Lunehead} & $<$29 & 5.6$\pm$1.1  & - & $<$0.6 \\ 
       \multicolumn{1}{ |c| }{ \textbf{Combined}} & \textbf{5.8$\pm$1.7} & \textbf{4.6$\pm$1.0}  & \textbf{267$\pm$9} & \textbf{$<$0.3} \\ 
    \hline\hline
         \multicolumn{1}{ |c| }{{Detector} } 
     & \rule{0pt}{2.4ex}{{\thtTtE\ }} 
     &  \rule{0pt}{2.4ex}{{\thtTtL\ }} 
     &  \rule{0pt}{2.4ex}{{\kfz\ }} 
     &  \rule{0pt}{2.4ex} {{\cosz\ }} \\
     \hline
        \multicolumn{1}{ |c| }{ Chaloner} & 1.9$\pm$1.0 & 1.4$\pm$0.3 & 29$\pm$6 & $<$0.3\\
        \multicolumn{1}{ |c| }{ Lunehead} & 2.7$\pm$1.1 & 1.8$\pm$0.4 & 28$\pm$6 & $<$0.4 \\
	\multicolumn{1}{ |c| }{ \textbf{Combined}} & \textbf{2.3$\pm$1.1} & \textbf{1.5$\pm$0.3}  & \textbf{29$\pm$6} & \textbf{$<$0.3} \\ 
	\hline
    \end{tabular}%
    }
  \label{tab:resistors}%
\end{table}%

For alumina, such as is used for the construction of the insulator in this model of resistor, the total neutron yield (due to the high cross-section ($\alpha$,n) reaction involving $^{27}$Al)  for the \utTe\ chain in equilibrium is calculated using the SOURCES4A software~\cite{sources4}. The software has been modified to extend the energies of $\alpha$-particles from the original energy cut of \SI{6.5}{\MeV} to \SI{10}{\MeV}~\cite{Carson2004667}, and to improve the cross-section library for a large number of materials~\cite{LEMRANI2006454,TOMASELLO2008431,TOMASELLO201070}, with newly added cross-sections calculated using the EMPIRE-2.19 code~\cite{HERMAN20072655}. In the worst case scenario, we assume that all the measured contamination in the surface mount resistors is confined to the ceramic. If this specific sample had been measured using Lunehead and equilibrium for all isotopes below \utTeL\ assumed then an overall neutron yield of \SI{1.6e-3}{neutrons \per year \per resistor} would have been reported. When the disequilibrium of \pbtoz\ is considered, this yield must be revised up by around 2.5$\times$. \textcolor{black}{This increase in neutron yield is due to the in-growth of its progeny \potoz\ which will reach secular equilibrium with \pbtoz\ after a period of about 2 years. This represents another potentially time dependent background in low-background experiments.}

\subsection{X-ray Fluorescence for Material Identification}
\begin{figure}
\centering
  \includegraphics[width=1.0\linewidth]{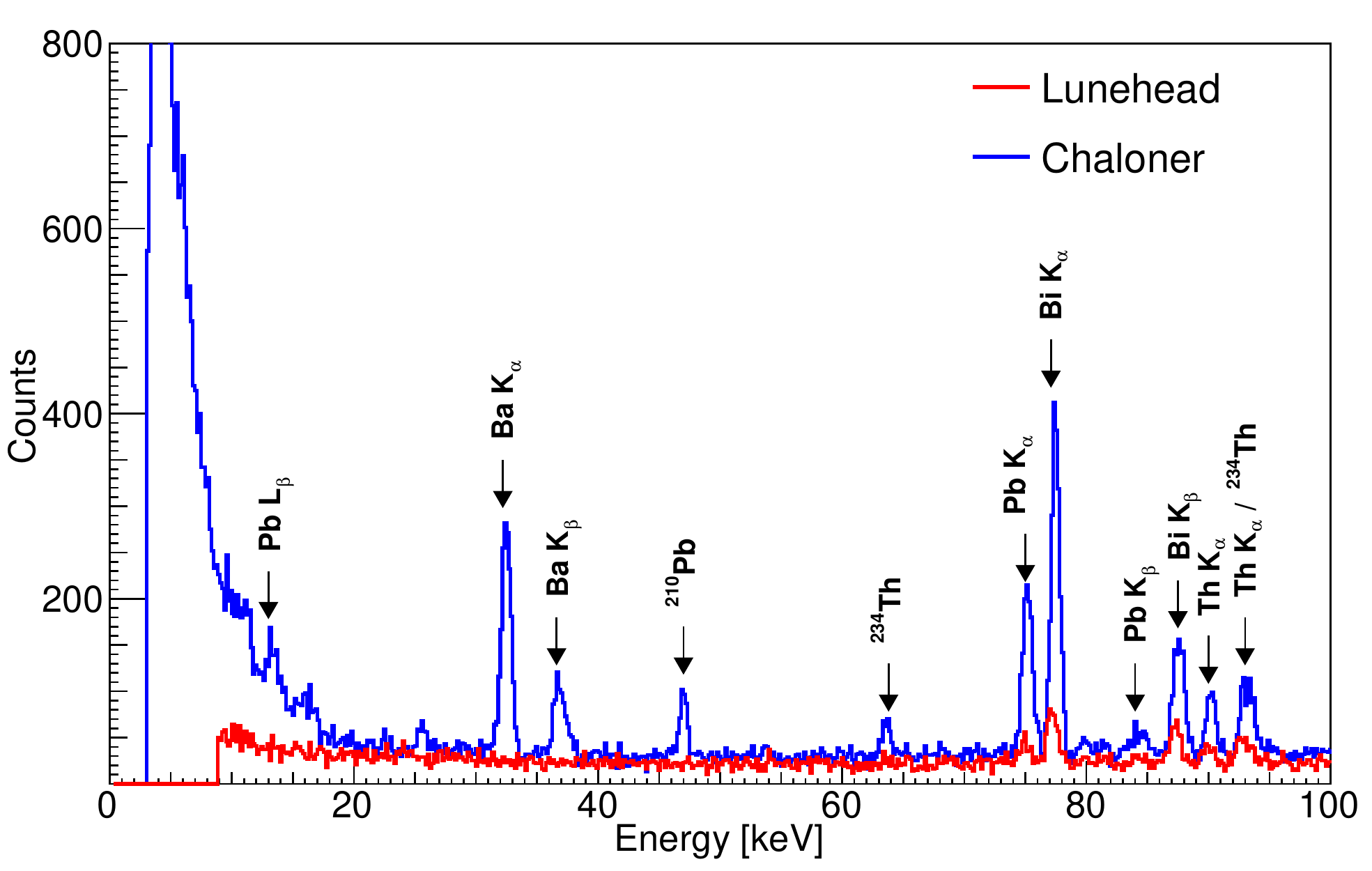}%
  \caption{Comparison between spectra from capacitor samples run on Lunehead (red) and Chaloner (blue). Chaloner retains sensitivities to low enough energies that we are able to identify barium K$_{\alpha}$ and K$_{\beta}$ X-rays at \SI{32}{\kilo\electronvolt} and \SI{36}{\kilo\electronvolt}, respectively.
\label{fig:capacitors}}
\end{figure}

The broad energy characteristics of Chaloner mean that sensitivity down to several keV is maintained. This allows for the identification of X-rays produced in X-ray fluorescence processes within a material. This has proven an interesting cross-check of dedicated elemental analysis in the identification of component material during the testing of capacitors. Figure~\ref{fig:capacitors} shows the low-energy spectrum of a capacitor sample. The increase in \pbtoz\ that was seen in the resistor sample is not observed in Chaloner but an array of X-ray peaks is present. The majority of these are lead and thorium X-rays which originate from the detector shielding but two peaks of additional interest are at \SI{32}{\kilo\electronvolt} and \SI{36}{\kilo\electronvolt}. The energies of these two peaks are consistent with the K$_{\alpha}$ and K$_{\beta}$ X-rays from barium, respectively.

\begin{figure}
\centering
  \includegraphics[width=1.0\linewidth]{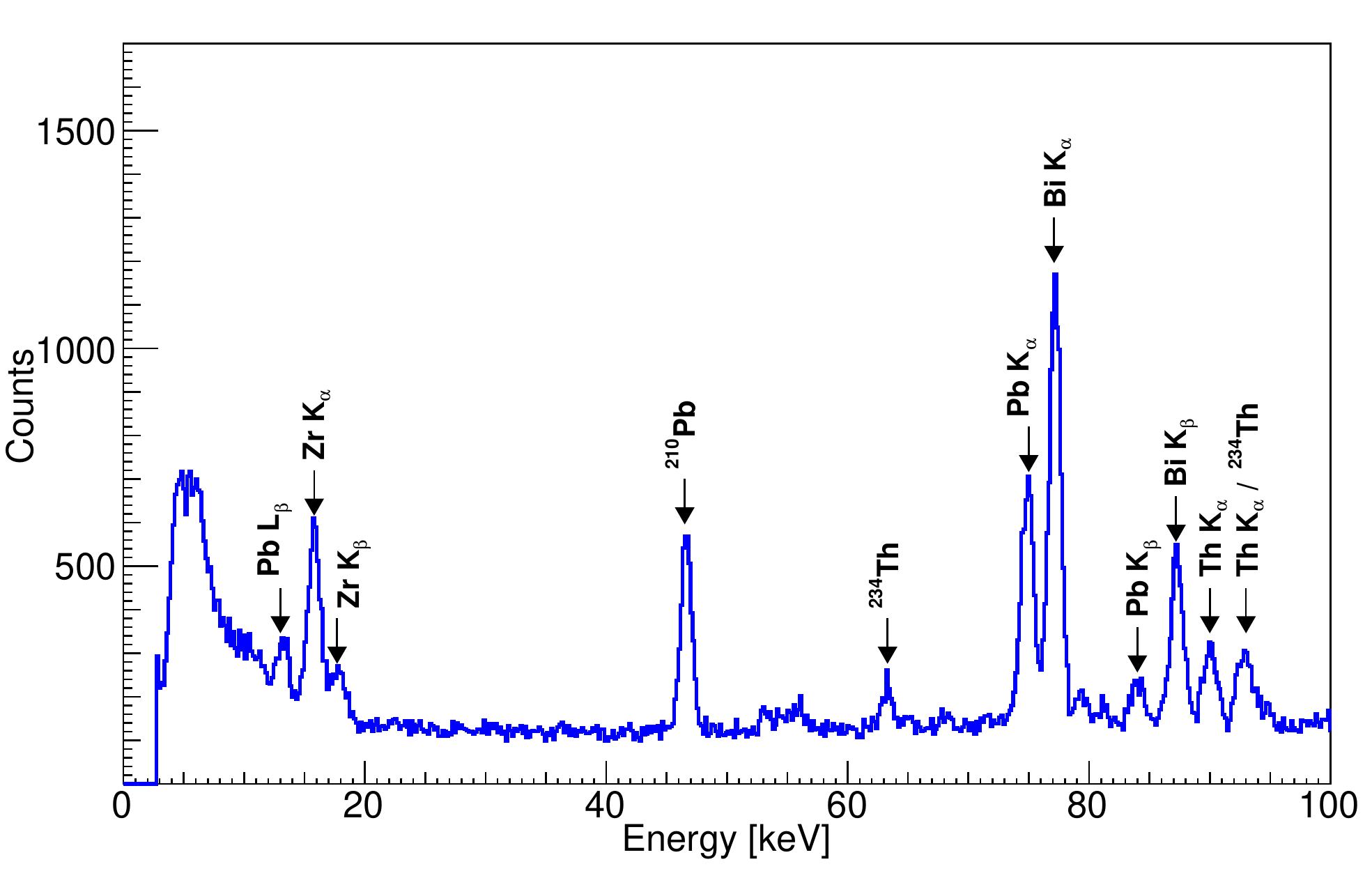}%
  \caption{Spectrum from a measurement of a ZrO$_{2}$ sample on Chaloner. The zirconium K$_{\alpha}$ line at \SI{16}{\kilo\electronvolt} is marked along with several other lines of interest.
\label{fig:zrspec}}
\end{figure}

The identification of these lines is consistent with results from dedicated scanning electron microscopy and energy dispersive X-ray elemental analysis showing that the dielectric used in the capacitors is barium titanate (BaTiO$_{3}$) rather than an alumina based ceramic. This material identification is of importance in the calculation of neutron yields for such materials with BaTiO$_{3}$ yielding some 8.8$\times$ lower value as compared to alumina for the same measured specific activities. As with the measurement of \pbtoz\ it would not be possible to perform such a material analysis using a coaxial detector alone. The rapidly increasing background seen in both Figures~\ref{fig:resistors} and~\ref{fig:capacitors} means that sensitivity is not maintained to X-rays from lower mass atoms for such low count rates. We are unlikely to be able to identify materials with major X-ray energies below that of the K$_{\alpha}$ X-ray of zirconium at \SI{16}{\kilo\electronvolt}. This X-ray is highlighted in the spectrum of a ZrO$_{2}$ sample in Figure~\ref{fig:zrspec}.

\section{Conclusions and Outlook}
\label{Sec::Conclusion}

The BUGS facility at the Boulby Underground Laboratory comprises three fully characterised low-background HPGe detectors, the sensitivity of which span energy ranges from \SI{30}{\kilo\electronvolt} to \SI{3000}{\kilo\electronvolt} which facilitates improved assay of early chain of \utTe, and assay of \pbtoz. The comprehensive characterisation of these detectors was performed using extended samples so as to represent the array of sample geometries that are screened for a typical low-background material screening programme. The MDAs calculated for each detector highlight the importance of maintaining high sensitivity to a broad range of \gray\ energies. We have shown that for low-density extended samples, we can achieve much higher sensitivity to the early chain of \utTe\ and can additionally assay \pbtoz\ using our BEGe-type detectors. This broad energy range has proven to be particularly important for samples where there is a large increase in levels of \pbtoz\ relative to the levels of \utTeL. Such a disequilibrium cannot be identified using coaxial detectors.

\begin{figure}
\centering
\subcaptionbox*{}{%
  \includegraphics[width=1.0\linewidth]{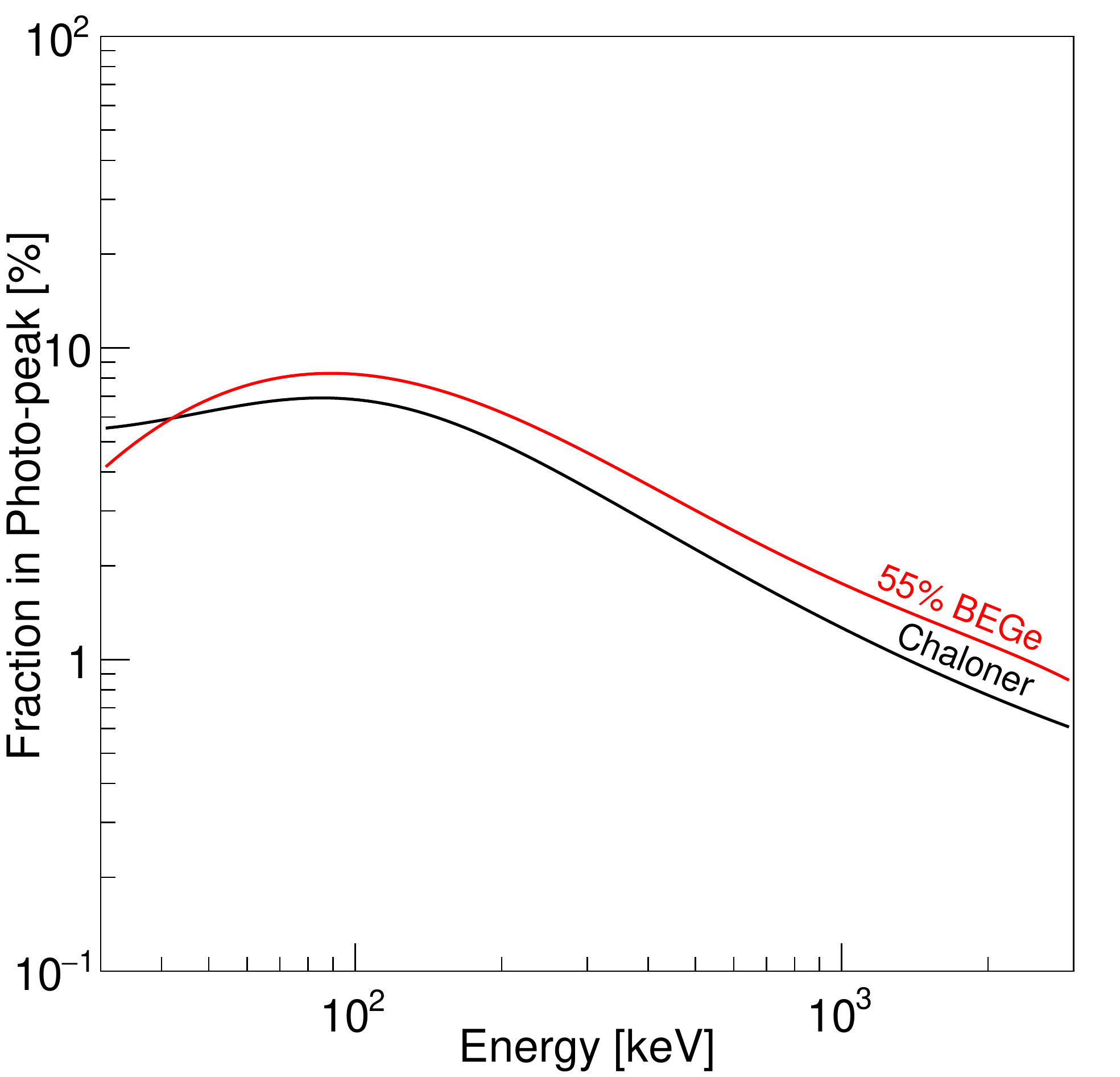}%
  
  }
\subcaptionbox*{}{%
  \includegraphics[width=1.0\linewidth]{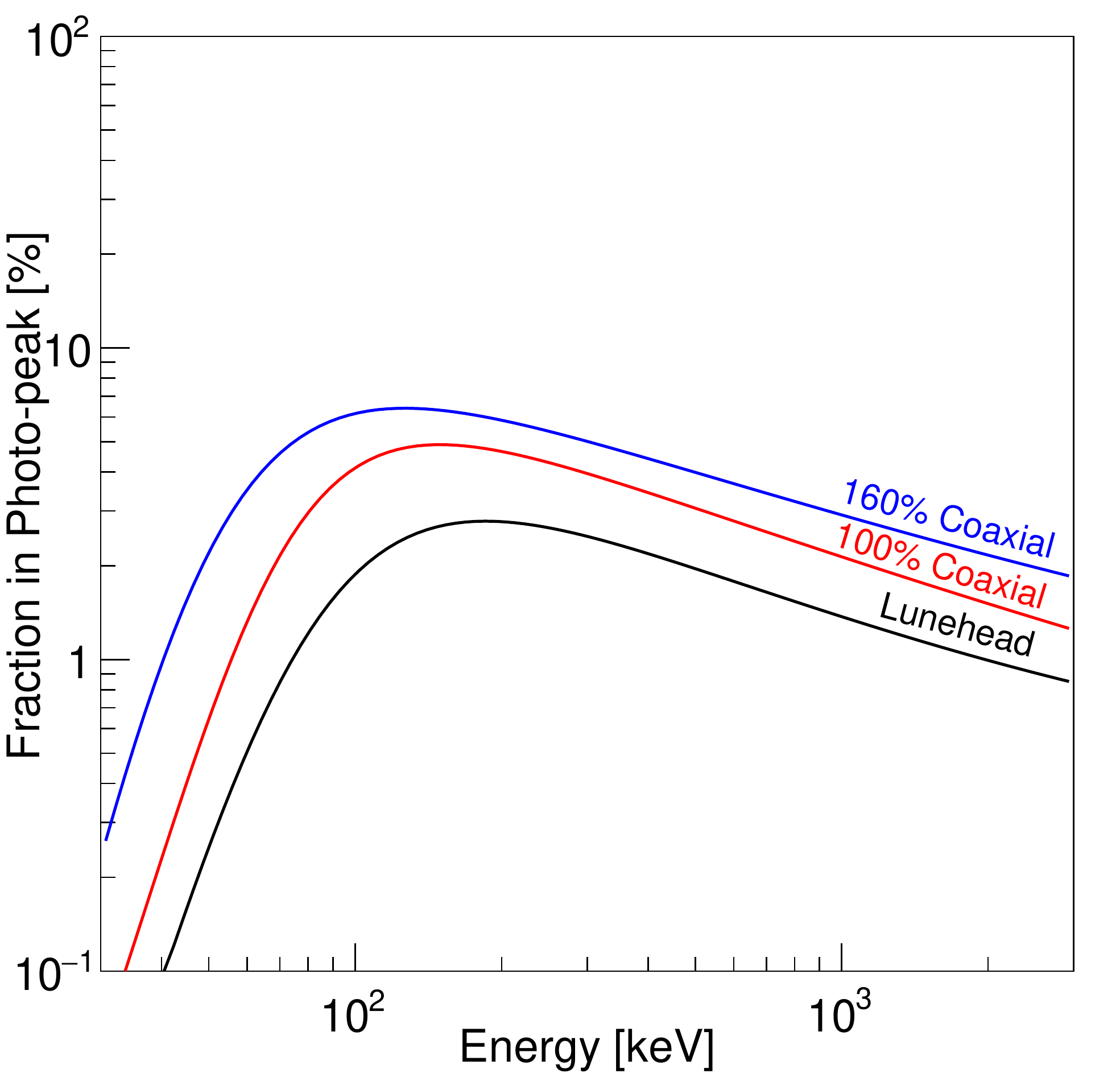}%
  }  
  \caption{Comparison between efficiencies for the \SI{850}{\gram} PTFE sample for (top) Chaloner and the future BUGS BEGe-type detector and (bottom) Lunehead and the two future BUGS p-type coaxial detectors. The new BEGe detector (red) has an all-aluminium cryostat. This leads to a sharper rolloff in efficiency at the lowest energies as compared to Chaloner (black). (bottom) The two new p-type detectors (red, blue) also display a slightly somewhat rolloff in efficiency as compared to Lunehead (black). These detectors also use all-aluminium cryostats rather than the carbon fibre end cap as used in Lunehead.
\label{fig:newDetectors}}
\end{figure}

BUGS is currently expanding to incorporate three new ultra-low-background HPGe detectors. These detectors, developed in collaboration with Mirion (Canberra), will comprise two coaxial HPGe detectors with nominal relative efficiencies of \SI{100}{\percent} and \SI{160}{\percent}, and a larger BEGe detector with a nominal relative efficiency of \SI{55}{\percent}. Material screening for the construction of these detectors has been performed using the BUGS detectors described here, and the low-background ICP-MS facility at University College London~\cite{dobson:inpress}. Figure~\ref{fig:newDetectors} shows the increase in efficiency that we would expect for detectors of this type in comparison with Lunehead, in the case of the new coaxial detectors, and Chaloner, in the case of the new BEGe-type detector. It is expected that these detectors will have intrinsic background of order 10$\times$ lower than that of our current detectors. 

The availability of these new detectors will significantly increase the sensitivity reach of BUGS, will allow rapid screening of samples suitable for current-generation low-background experiments, and will facilitate the screening of materials to the level needed for next-generation low-background experiments, particularly the so-called `Generation-3' dark matter experiments beyond LZ, and next-generation \dbd\ searches.

\section{Acknowledgements}
This work was supported by the U.K. Science \& Technology Facilities Council under award numbers ST/L003228/1, ST/K006428/1, ST/M003655/1, ST/M003981/1, ST/M003744/1, ST/M003639/1, ST/M003604/1, and ST/M003469/1. University College London thanks the U.K. Royal Society for travel funds under the International Exchange Scheme (IE141517). The authors would like to thank Israel Chemicals Ltd UK (ICL-UK) for their support in the construction and operation of the new Boulby Underground Laboratory and the BUGS facility. We would also like to acknowledge Lawrence Berkeley National Laboratory, particularly Kevin Lesko, Al Smith, Andrew Cole and Keenan Thomas for their help in confirming the characterisation of these detectors. We wish to thank Richard Howard, Phil Callow, and Stan Mahony from Lead Shield Engineering Ltd. for their help in the design, manufacture and installation of the shielding castles. We wish to thank Pascal Quirin and Tony Francis for their continuing help in the development of detectors for BUGS. Finally, we wish to thank our UK colleagues from the LZ and SuperNEMO collaborations for the initial and ongoing support of this facility.

\section{Bibliography}
\bibliographystyle{hunsrt}
\bibliography{bibliography}

\end{document}